\documentclass[aps,pra,twoside,twocolumn,amsfont,amssymb,amsmath,showpacs,preprintnumbers,longbibliography]{revtex4-2}
\usepackage{breqn}
\usepackage{etoolbox}
\usepackage{amsmath}
\usepackage{bm}
\usepackage{color}
\usepackage{dcolumn}
\usepackage{float}
\usepackage{graphicx}
\usepackage{epstopdf}
\usepackage[colorlinks,linkcolor=red,citecolor=blue]{hyperref}
\usepackage{mathbbol}
\usepackage{mathrsfs}
\usepackage{placeins}
\usepackage{verbatim}
\usepackage{url}
\usepackage{breakurl}

\newcommand{\CAP}{{\scriptscriptstyle\mathrm{CAP}}}
\newcommand{\XUV}{{\scriptscriptstyle\mathrm{XUV}}}

\newcommand{\IR}{{\scriptscriptstyle\mathrm{IR}}}

\begin{document}

\title{Ionic Coherence in Resonant Above-Threshold Attosecond Ionization Spectroscopy}

\author{Saad Mehmood$^1$,  Eva Lindroth$^2$, and Luca Argenti$^{1,3}$}\email{luca.argenti@ucf.edu} 
\affiliation{$^1$Department of Physics University of Central Florida}
\affiliation{$^2$Department of Physics Stockholm University Stockholm}
\affiliation{$^3$CREOL University of Central Florida Orlando Florida}


\begin{abstract}
The ionization of atoms with sequences of attosecond pulses gives rise to excited ionic states that are entangled with the emitted photoelectron. Still, the ionic ensemble preserves some coherence that can be controlled through the laser parameters. In helium, control of the $2s/2p$ He$^+$ coherence is mediated by the autoionizing states below the $N=2$ threshold [Phys. Rev. Res. {\bf 3}, 023233 (2021)]. In the present work we study the role of the resonances both below and above the $N=3$ threshold on the coherence of the $N=3$ He$^+$ ion, in the attosecond pump-probe ionization of the helium atom, which
we simulate using the newstock \emph{ab initio} code. Due to the fine-structure splitting of the N=3 He$^+$ level, the ionic dipole beats on a picosecond timescale.  We show how, from the dipole beating, it is possible to reconstruct the polarization of the ion at its inception.
\end{abstract}

\maketitle

\section{Introduction}

The typical spectrum of a poly-electronic atom comprises several excited bound and mestastable states. In the vicinity of an ionization threshold, many of these states can be coherently populated with short pulses, giving rise to electronic motion that unfolds on a sub-femtosecond timescale~\cite{Krausz2009,RevModPhys.87.765,Calegari_2016}. The advent of table-top sources of attosecond light pulses has unlocked the door to the observation and control of such electronic motion on its natural time scale~\cite{Krausz2009,Lepine2013,Lepine2014,Leone2014,LeoneNeumark2016,Nisoli2017,Sansone2012,PazourekRMP2015}. Once the production of attosecond pulses became routine, attosecond XUV-pump IR-probe photoelectron spectroscopies emerged as a powerful tool to explore attosecond dynamics at small scales~\cite{Ciappina_2017,DombiNano_2020} and charge-transfer processes in molecular systems~\cite{Haessler2010,Goulielmakis2008, Itatani2002, Villoresi2006,Horvath2007}. The generation of coherent superpositions of electronic states above the ionization threshold bears the promise of quantum control in the electronic continuum. However, since ions and photoelectrons normally form entangled pairs, either of the photofragments is only partially coherent, and their coherence depends on the parameters of the ionization process~\cite{SFI_Pabst,DecoherencePRX}. 
In a recent work, we examined how the autoionizing states below the $N=2$ threshold of the helium atom could be leveraged to control the relative coherence between the $2s$ and $2p$ state of the He$^+$ ion~\cite{Mehmood2021}. In this work we explore the entangled character of the wavefunction for system formed by a He$^+$ parent ion in the $N=3$ manifold and by the photoelectron that emerge from the ionization of the helium atom. We also describe a protocol to reconstruct the density matrix of the He$^+$ ensemble from the measurement of the picosecond time-scale beatings of its dipole moment, which are caused by the fine splitting of the $N=3$ level.

On a femtosecond timescale, the effects of spin-orbit interactions do not manifest themselves, and hence the coherence between He$^+$ ionic states with the same principal quantum number appears as a permanent polarization of the ion. The ion can be produced in such a polarized state by triggering ionization in the presence of a strong infrared dressing pulse with a pulse of extreme ultraviolet radiation with duration shorter than half the period of the IR pulse~\cite{Ossiander2016,GoulielmakisNAT2010}. Alternatively, it is possible to exploit the interference between different resonant ionization paths~\cite{Mehmood2021,Guillemin2015,ShapiroPRA2011}.
Resonant states play a crucial role in multiphoton ionization since, in contrast to photofragments, they are fully coherent states and they have long lifetime compared with direct-ionization photoelectron wavepackets, which leave the interaction region in a matter of few tens attoseconds~\cite{Argenti2010,ArgentiScience,EvaLuca_ChemRev,LucaPRL2014,Kotur2016,Ott2014}. Such long-lived metastable states provide resonant multi-photon pathways for ionization in atoms~\cite{Nagasono2007,Htten2018}, ultrafast decay of electrons~\cite{Fohlisch2005} and dissociative photoionization in molecules~\cite{Sansone2010, Martin2007}. A variety of experiments have studied autoionizing resonances in an attempt to resolve the electron-correlation driven dynamics that underpins Auger decay~\cite{Doughty2011,Wickenhauser2006,Cirelli2018,ArgentiScience,Ott2014}.
In the present study, we examine the interplay between the autoionizing states below and above the $N=3$ threshold in multiphoton ionization paths that are resonant both in the intermediate and in the final states. We do so by comparing the relative coherences of the He$^+$ $3s$, $3p_m$, and $3d_m$ states computed either in absence or in the presence of the resonances above the N=3 threshold. 
As in the case of the $N=2$ manifold~\cite{Mehmood2021}, in the non-relativistic limit, $N=3$ He$^+$ parent-ion states are degenerate, and hence their coherence results in a permanent dipole moment. On a timescale of few picosecond, the dipole moment fluctuates even in absence of external fields, due to fine-structure terms, and in particular to spin-orbit interaction~\cite{bookSpringer}. In contrast to the $N=2$ case, however, the reconstruction of the He$^+$ density matrix from the sole observation of the system's dipole fluctuations cannot be complete. Here we examine a general algorithm which gives the subset of initial coherences that are constrained by the observation of the ionic dipole alone.

The paper is organized as follows.  Section~\ref{sec:TheoreticalMethods} offers an overview of the \emph{ab initio} theoretical and numerical methods used to compute, at the end of a pulse sequence, the photoelectron distribution entangled with each ion.
Section~\ref{sec:Simulations} describes the pump-probe setup used for the simulations, it  discusses the partial photoelectron distributions as well as the corresponding reduced density matrix for the ion. Section~\ref{sec:Reconstruction} describes the reconstruction of the ionic coherence phase from the picosecond beating of the ion dipole. Finally, Section~\ref{sec:Conclusions} summarizes the conclusions and perspectives of this work.

\section{Theoretical Methods}\label{sec:TheoreticalMethods}
\color{black}
In this work, we describe the bound and single-ionization states of the helium atom using the close-coupling approach~\cite{Argenti2006,Argenti2010,Carette2013}.
The theory and most recent implementation of time-dependent close-coupling code dedicated to the helium atom and for the NewStock code, for general polyelectronic atoms, has been described elsewhere~\cite{Argenti2021,Mehmood2021,Cariker2021}. Here we offer a brief summary of the protocols implemented in NewStock, adapted to the helium atom, which were used for the present calculations. The close-coupling (CC) representation of a single ionization function $\Psi(x_1,x_2;t)$ for the helium atom is,
\begin{eqnarray}
\Psi(x_1,x_2;t)&=&\frac{1-\hat{\mathcal{P}}_{12}}{\sqrt{2}}\sum_{\Gamma\alpha}\Phi^{\Gamma}_\alpha(x_1;\hat{r}_2,\zeta_2)\varphi^\Gamma_\alpha(r_2;t) +\nonumber\\
&+&\sum_{\Gamma i}\chi^\Gamma_i(x_1,x_2) c^\Gamma_i(t).\label{eq:tdCC}
\end{eqnarray}
where $x_i=(\vec{r}_i,\zeta_i)$ is the spatial-spin component of the i-th electron, $\hat{\mathcal{P}}$ swaps the coordinates of the two electrons, $\Phi^{\Gamma}_\alpha(x_1;\hat{r}_2,\zeta_2)$ is a channel function in which the orbital and spin angular momentum of the ion are coupled to the photoelectron's,
\begin{equation}\label{eq:CCchannel}
\Phi^{\Gamma}_\alpha(x_1;\hat{r}_2,\zeta_2)=R_{N_\alpha L_\alpha}(r_1) \mathcal{Y}_{L_\alpha \ell_\alpha}^{LM}(\hat{r}_1,\hat{r}_2)\Theta_{S\Sigma}(\zeta_1,\zeta_2),
\end{equation}
and $\varphi^\Gamma_\alpha(r_2;t)$ is the radial component of the photoelectron wave function in channel $(\Gamma,\alpha)$~\cite{LucaAlvaroPRA2015,Ott2014,LucaPRL2014}.
In~\eqref{eq:CCchannel}, $\mathcal{Y}_{ab}^{c\gamma}$ and $\Theta_{S\Sigma}$ are  bipolar spherical harmonics and two-electron spin functions, respectively,
\begin{equation}
\begin{split}
&\mathcal{Y}_{\ell_1\ell_2}^{LM}(\hat{r}_1,\hat{r}_2)=\hspace{-5pt}\sum_{m_1m_2}C_{\ell_1 m_1,\ell_2 m_2}^{LM} Y_{\ell_1 m_1}(\hat{r}_1)Y_{\ell_2 m_2}(\hat{r}_2),\\
&\Theta_{S\Sigma}(\zeta_1,\zeta_2)=\sum_{\sigma_1\sigma_2}C_{\frac{1}{2}\sigma_1,\frac{1}{2}\sigma_2}^{S\Sigma}{^2\chi_{\sigma_1}(\zeta_1)} {^2\chi_{\sigma_2}(\zeta_2)}.
\end{split}
\end{equation}
The collective symmetry label $\Gamma$ stands for the quantum numbers of the two-electron system, i.e., the parity $\Pi$ and the total orbital and spin angular momenta and projections $L$, $S$, $\Sigma$, and $M$, whereas $\alpha$ identifies the parent-ion shell, $N_\alpha L_\alpha$ and the photoelectron orbital angular momentum $\ell_\alpha$.  Finally, the functions $\chi_i(x_1,x_2)$ are symmetry adapted 2-electron configurations state functions (CSF) $^{2S+1}(n_1\ell_1,n_2\ell_2)_{LM}$ with principal quantum numbers $n_i$ and angular quantum numbers $\ell_i$ restricted to $n_i\leq N_{\mathrm{max}}$, $\ell_1\leq L_{\mathrm{max}}$.
The CC wave functions~\eqref{eq:tdCC}, and the time evolution of helium from an initial bound state as a result of the interaction with external fields are computed with \verb|newstock|~\cite{Carette2013,Argenti2021}. In this work, the time dependent Hamiltonian $H(t)$ comprises the electrostatic Hamiltonian $H_0$ and the velocity-gauge interaction Hamiltonian $H_I$,
\begin{equation}
\begin{split}
H(t)&=H_0 + H_I(t)\\
H_0 &= \frac{p_1^2+p_2^2}{2}-\frac{2}{r_1}-\frac{2}{r_2}+\frac{1}{r_{12}}\\
H_I(t) &= \alpha \vec{A}(t)\cdot(\vec{p}_1+\vec{p}_2),
\end{split}
\end{equation}
where $\vec{A}(t)$ is the vector potential and $\alpha$ is the fine-structure constant, $\alpha = e^2/\hbar c\approx 1/137$~\cite{CODdataNIST}. Unless stated otherwise, atomic units and the Gauss System are used. The reduced radial function of all one-electron orbitals, $r\varphi(r)$, is expanded in B-splines~\cite{deBoor1978,Bachau2001,Argenti2006}. 
In this work, two CSF basis are used. A first smaller set, referred to as case 1, comprises all the configurations of the form $n\ell i\ell'$ with $n\leq 3$, $\ell \leq 2$, $\ell'\leq 5$, and the index $i$ runs over all the radial functions in the quantization box (several hundred), which do not necessarily resemble hydrogenic bound orbitals. This first basis is unable to represent any resonance above the $N=3$ threshold, as these are known to originate from configurations of the form $n\ell n'\ell'$, with $n\geq 4$~\cite{Rost_1997}. A second larger basis, referred to as case 2, comprises all the configurations of the form $n\ell i\ell'$ with $n\leq 4$, $\ell \leq 3$,  $\ell'\leq 5$, and the index $i$ has the same meaning as above. This second basis does give rise to resonances between the $N=3$ and $N=4$ thresholds, most of which have dominant configuration $4\ell n'\ell'$. This energy interval includes also a couple of so-called intruder states of the form $5\ell 5\ell'$~\cite{Burgers1995,Rost_1997,Argenti2006}. These intruder states, however, fall close to the $N=4$ threshold, which is not reached by our pump-probe scheme, and hence are not expected to play any major role here. Since the He$^+$ ion has only one electron, within the electrostatic approximation, its eigenstates, numerically computed using the ATSP2K package~\cite{Froese2007}, are virtually exact. 
In the present work we will focus our attention on the effects of the probe pulse to the lowest perturbative order. For this reason, for both cases, we will consider only the total symmetries $^1$S$^e$, $^1$P$^o$, and $^1$D$^e$, using $^1$F$^o$ only to check the convergence of our simulations. Notice that we are not including any $^1$P$^e$, $^1$D$^o$, or $^1$F$^e$ state in the basis because non-natural symmetries (i.e., whose parity differs from the total angular momentum parity) cannot be populated by collinearly polarized light pulses starting from an initial state with magnetic quantum number $M=0$.
Each symmetric space comprises a localized channel constructed by adding, in all possible ways, an electron to a CSF configuration in any of the $3s$, $3p$, $3d$, $4s$, and $4p$ active orbitals. There are 20 (10), 20 (8), 21 (7) and 12(2) such states for the  $^1$S$^e$, $^1$P$^o$, $^1$D$^e$ and $^1$F$^o$ symmetries for the case-2 (case-1) basis, respectively.
To represent the radial part of both bound and continuum atomic orbitals, we use B-splines of order $7$ with asymptotic separation between consecutive nodes of $0.4$~a.u., up to a maximum radius of $300$~a.u. With this choice, each PWC comprises approximately 1285 states. 
The \verb|newstock| package builds the field-free hamiltonian matrix $\mathbf{H}_{ij}^\Gamma = \langle \Psi^\Gamma_i | \hat{H}_0  | \Psi^\Gamma_j \rangle$ for each of the four main symmetries $\Gamma = ^1$S$^e$, $^1$P$^o$, $^1$D$^e$, $^1$F$^o$ where $\Psi^\Gamma_j$ is any of the functions in the generalized close-coupling space with symmetry $\Gamma$, as well as the reduced dipole matrix elements  $\langle \Psi^\Gamma_i \| P_1 \| \Psi^{\Gamma'}_j \rangle$ between S and P, P and D and between D and F states. 
The initial ground state is obtained by diagonalizing the Hamiltonian in the $^1$S$^e$ sector of the full close-coupling space.  The time evolution of the atomic wave function, from the initial ground state, under the influence of the external pulses is governed by the time-dependent Schr\"odinger equation (TDSE), 
\begin{equation}
    i\partial_t\Psi(t) = H(t) \Psi(t),
\end{equation}
which is integrated numerically in time steps of the duration of approximately $dt=0.033$~a.u., using a unitary second-order exponential split operator, with the inclusion of a complex absorption potential ($\CAP$) in the last 50~Bohr Radii of the spherical quantization box, to prevent unphysical reflections from the box walls,
\begin{equation}
\begin{split}
    \Psi(t+dt) &= U_{\CAP}(dt)\,U(t+dt,t) \Psi(t_0)\\
    U(t+dt,t)&=e^{-i\,H_0\,dt/2}e^{-i\,H_I(t+dt/2) \,dt}e^{-i\,H_0 \,dt/2}\\
    U_{\CAP}(dt) &= e^{-iV_{\CAP} dt}\\
    V_{\CAP}&=-ic\sum_{i=1}^2\theta(r_i-R_{\CAP})(r_i-R_{\CAP})^2,\\ R_\CAP&=250~\mathrm{a.u.},
\end{split}
\end{equation}
where $\theta(x)$ is the Heaviside step function ($\theta(x)=0$ if $x<0$, $\theta(x)=1$ if $x>0$), and $c$ is a positive real parameter, $c=5\cdot 10^{-4}$. The reflection by the CAP itself and by the box boundary are both negligible. The unitary evolution under the action of the dipole operator is evaluated with an iterative Krylov method. 

The photoelectron component of the wavefunction grows rapidly in size. Indeed, in 100~fs, the typical size of an attosecond pump-probe time-delay scan, a photoelectron with asymptotic energy of about 1~a.u. ($\simeq$\,27~eV) covers a distance of more than 4000 Bohr radii. In order to recover the spectrum of the photoelectron, the whole wavefunction must be preserved. On the other hand, once the photoelectron has reached a distance of few hundred atomic units, the coupling between different channels is negligible and hence the electron propagates in what essentially is a pure monopolar electrostatic potential plus the potential due to the external radiation field. In these conditions, an explicit numerical representation of the whole photoelectron wavepacket is highly impractical. Luckily, at large distances, we can exploit a useful approximation, which we describe below. The propagator from an initial time $t_0$ to a final time $t$, reads
\begin{equation}
    U(t,t_0)=\hat{T}\exp\left[-i \int_{t_0}^t dt_1 H(t_1)\right],
\end{equation}
where $\hat{T}\exp$ is the time-ordered exponential~\cite{sakurai2017modern}. This propagator can be factorized using Magnus expansion~\cite{Blanes2009},
\begin{equation}
    U(t,t_0)=\exp\left[-i\Omega(t,t_0)\right],
\end{equation}
where
\begin{equation}
\begin{split}
    \Omega(t,t_0)&=\sum_{n=1}^\infty\Omega_n(t,t_0),\\
    \Omega_1(t,t_0)&=\int_{t_0}^t H(t_1) dt_1\\
    \Omega_2(t,t_0)&=\frac{1}{2i}\int_{t_0}^t dt_1\int_{t_0}^{t_1}dt_2\,[H(t_1),H(t_2)]\\
    \cdots
\end{split}
\end{equation}
When the photoelectron is sufficiently far from the associated ion, the time evolution of the electron and that of the ion become independent. The Hamiltonian of the photoelectron, in particular, becomes
\begin{equation}
    H(t)=\frac{p^2}{2}-\frac{1}{r}+\alpha\vec{A}(t)\cdot\vec{p}.
\end{equation}
The commutator between the field-free component of the photoelectron Hamiltonian and the interaction term scales as $1/r^2$, and hence, for a photoelectron sufficiently far from the ion, all higher order terms in the Magnus expansion become negligible,
\begin{equation}
\begin{split}
U(t,t_0)&\approx\exp\left[-i\int_{t_0}^t H(t)dt\right]=\\
&=\exp\left[-i \left(\frac{p^2}{2}-\frac{1}{r}\right)(t-t_0)-i\vec{p}\cdot\int_{t_0}^t\vec{A}(t)dt\right].
\end{split}
\end{equation}

Whenever the integral of the vector potential between an initial and a final time vanishes, therefore, the interaction term itself vanishes, and hence the photoelectron propagator can be accurately approximated by the field-free propagator alone. We take advantage of this circumstance in our simulations because,
about every half period of the MIR field, $t_i$, the integral $\int_{t_i}^\infty A(t') dt'$ does vanish. At these times, we split the wavefunction into a mid/short-range ($r\lesssim 150$~a.u.) and a mid/long-range component ($r\gtrsim 150$~a.u.), 
\begin{equation}
\begin{split}
\Psi_{0,\textsc{SR}}(t_0)&=\Psi(t_0)\\
\Psi_i(t_i) &= \Psi_{i,\textsc{SR}}(t_i) + \Psi_{i,\textsc{LR}}(t_i),\\
\Psi_{i,\textsc{SR}}(t_i) &= P |\Psi_{i-1,\textsc{SR}}(t_i)\rangle,\\
\Psi_{i,\textsc{SR}}(t_i) &= (1-P) |\Psi_{i-1,\textsc{LR}}(t_i)\rangle
\end{split}
\end{equation}
where $P$ extracts the short-range distance of the wavefunction,
\begin{equation}
P = \sum_i \Phi\left(\frac{r_i-R_{\textsc{MASK}}}{\sigma_{\textsc{MASK}}}\right),
\end{equation}
and $\Phi(x)$ is the normal cumulative distribution function, $\Phi(x)=(2\pi)^{-1/2}\int_{-\infty}^x e^{-t^2/2}dt$. By choosing $R_{\textsc{MASK}}=R_{\textsc{BOX}}/2$, and $1\ll\sigma_{\textsc{MASK}}\ll R_{\textsc{MASK}}$, we ensure that $\Psi_{\textsc{LR}}(t_i)$ is non negligible only at large distances from the ion, that it does not acquire unphysical incoming components (and hence, that under the subsequent evolution it remains at large distance from the ion), and that $\Psi_{\textsc{SR}}(t_i)$ is negligible in the last third or so of the quantization box.

As explained above, as far as the wavefunction long-range component extracted at the times $t_i$ are concerned, the field-free and the interaction Hamiltonians effectively commute, and the effect of the interaction Hamiltonian itself vanishes. Each of these long-range components, therefore, can be propagated analytically to the end of the pulse, $t$, by means of the field-free evolution operator,
\begin{equation}
\Psi_{i,\textsc{LR}}(t)=e^{-iH_0(t-t_i)}\Psi_{i,\textsc{LR}}(t_i).
\end{equation}
Furthermore, once the external pulses are over, the last residual short-range component $\Psi_{N,\textsc{SR}}(t_N)$ can also be propagated in the same way,
\begin{equation}
\Psi_{N,\textsc{SR}}(t)=e^{-iH_0(t-t_N)}\Psi_{N,\textsc{SR}}(t_N).
\end{equation}
This procedure allows us to carry out simulations for long pulses in comparatively small quantization boxes. These propagation can be carried out exactly, since in our approach we operate in a spectral basis.

We determine, in interaction representation, the probability amplitude  $\mathcal{A}_{a;\varepsilon\hat{\Omega}\sigma}$ for detecting in coincidence the parent ion in the oriented ($M_a$, $\Sigma_a$) state $a$ and the photoelectron with energy $\varepsilon$, along the direction $\hat{\Omega}$, and with spin projection $\sigma$, by projecting the full wavepacket on a complete set of scattering states, 
\begin{equation} 
\mathcal{A}_{a;\varepsilon\hat{\Omega}\sigma} = e^{i(E_a+\varepsilon)t} \langle \Psi_{a;\varepsilon\hat{\Omega}\sigma}^- | \Psi(t)\rangle,
\end{equation}
where $|\Psi_{a;E\hat{\Omega}\sigma}^-\rangle$, normalized as $\langle \Psi_{a;E\hat{\Omega}\sigma}^-|\Psi_{b;E'\hat{\Omega}'\sigma'}^-\rangle = \delta_{ab}\delta^{(2)}(\hat{\Omega}-\hat{\Omega}')\delta_{\sigma\sigma'}$, is a scattering state in which the parent ion and the photoelectron are not angularly or spin coupled, which fulfills incoming boundary conditions, and in which the photoelectron has a well defined propagation direction $\hat{\Omega}$~\cite{Argenti2010,ArgentiPRA2013,Argenti2006R}. We can now use the partitioning of the wavefunction in a last short-range component and many long-range components,
\begin{equation}
    \Psi(t) = \Psi_{N,\textsc{SR}}(t) + \sum_{i=1}^N \Psi_{i,\textsc{LR}}(t),
\end{equation}
to compute the amplitude as
\begin{equation} 
\begin{split}
\mathcal{A}_{a;\varepsilon\hat{\Omega}\sigma} &= e^{i(E_a+\varepsilon)t_N} \langle \Psi_{a;\varepsilon\hat{\Omega}\sigma}^- | \Psi_{N,\textsc{SR}}(t_N)\rangle + \\
&+\sum_{i=1}^N e^{i(E_a+\varepsilon)t_i}\langle \Psi_{a;\varepsilon\hat{\Omega}\sigma}^- | \Psi_{i,\textsc{LR}}(t_i)\rangle.
\end{split}
\end{equation}

The reduced density matrix for the parent-ion ensemble generated in a simulation with a pump-probe delay $\tau$, $\rho_{\alpha\beta}(\tau)$, is obtained tracing out the photoelectron states~\cite{FanoDM_1957},
\begin{equation}
\rho_{\alpha\beta}(\tau) = \sum_{\sigma}\int d^{3}k  \mathcal{A}_{\alpha\vec{k}\sigma}(\tau)\mathcal{A}^*_{\beta\vec{k}\sigma}(\tau).
\end{equation}
We define the coherence between ionic states~\cite{FanoDM_1957,Pabst2011} as 
\begin{equation}\label{eq:coherence}
    g_{\alpha\beta}(\tau) = \frac{|\rho_{\alpha\beta}(\tau)|}{\sqrt{\rho_{\alpha\alpha}(\tau)\rho_{\beta\beta}(\tau)}}.
\end{equation}
When considering ionic states with the same principal quantum number, within the electrostatic approximation, the density matrix $\rho_{\alpha\beta}(\tau)$ is independent of time even in the Schr\"odinger representation.

Due to the fine-structure terms in the Hamiltonian, $H_{\mathrm{fs}}$, however, even the block of the density matrix with a same principal quantum number undergoes slow periodic oscillations, on a picosecond timescale, reproduced by the unitary transformation
\begin{equation}\label{eq:TDDM}
\rho(t;\tau)=e^{-iH_{fs}t}\rho(\tau)e^{iH_{fs}t}.
\end{equation}
By the same token, the ion dipole moment is not stationary either, exhibiting fluctuations at the Bohr fine frequencies of the ion, $ \langle \mu_z(t;\tau)\rangle = \mathrm{Tr}[\mu_z\rho(t;\tau)]$. The details of the transformation to the basis that diagonalize the fine-structure Hamiltonian and the consequent time evolution of the He$^+$ state in the $N=3$ manifold is discussed more in detail in Sec.~\ref{sec:Results}.

\section{Simulation Scheme}\label{sec:Simulations}
\begin{figure}[hbtp!]
\begin{center}
    \includegraphics[width=0.8\columnwidth]{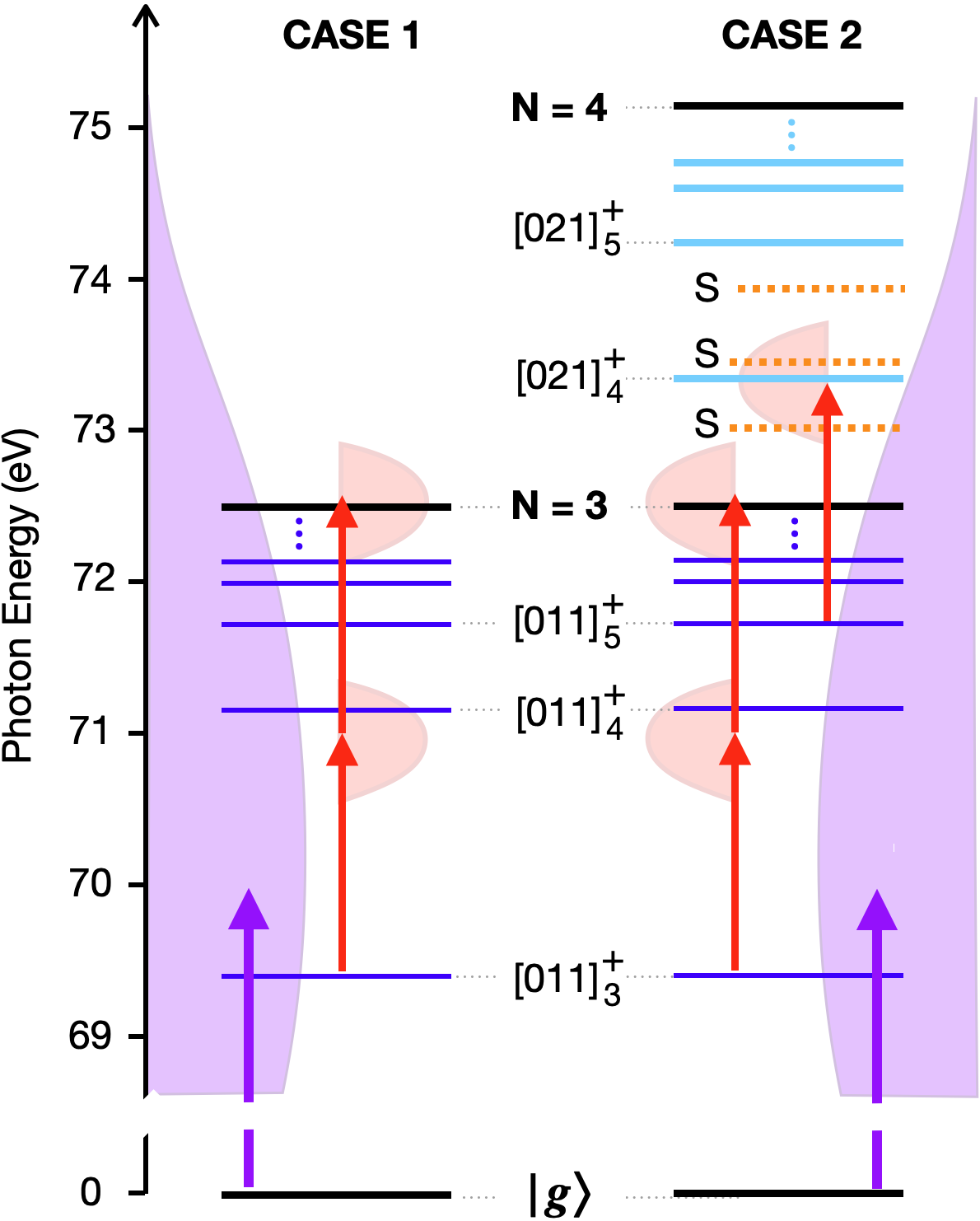}
    \caption{Energy Scheme for the two cases is compared: (case 1) Only the resonances that lead up to $N=3$ ionization threshold are included. An XUV, centered just above the [011]$_{3}^+$ resonance can coherently populate many close by resonances, along with direct ionization to $N_L^{'} \epsilon_\ell^{'}$ channels. In (case 2), we have all that and additional resonances above $N=3$ that we can coherently excite with the rest of the system. These additional S and P resonances will interfere with the resonance below the threshold to modify the interference pattern which can be probed the time delay scan. }
    \label{fig:EnergyScheme}
\end{center}
\end{figure}

Figure~\ref{fig:EnergyScheme} illustrates the energy scheme of the system, in relation to the transitions above the $N=3$ threshold promoted by the XUV-pump and the IR-probe pulses, for the two cases examined in this work, i.e., excluding (case 1) or including (case 2) the $N=4$ close-coupling channels.
The comparison between these two cases allows us to highlight the role of the Feshbach resonances above threshold in influencing the polarization of the $N=3$ ion.

A weak single attosecond XUV pulse excites the neutral helium atom from the ground state to the $^1$P$^o$ continuum in the energy interval between the first  autoionizing states in the series converging to the $N=3$ threshold up to the $N=4$ threshold. 
In case 1, the configuration basis does not give rise to any autoionizing state above $N=3$, whereas case 2 features several resonances from the series converging to $N=4$, starting from $\sim$0.5~eV above the $N=3$ threshold.

As discussed in~\cite{Saad2021}, where ionization in proximity of the $N=2$ threshold was considered, a single-photon transition cannot give rise to an asymmetrically polarized ion. Instead, multiphoton transitions are necessary to induce and control any coherence in the $N=2$ He$^+$ ion. The case of the ionization above the $N=3$ threshold, examined here, is different in that the $3s$ and the $3d$ states have the same parity, and hence even one-photo transitions can result in the formation of a partly coherent state. Even in this case, however, the parent ion is not electrically polarized. 

To polarize the parent ion, it is still necessary to associate the XUV pulse with additional control fields. In our simulation, we include an $800$~nm IR-probe pulse, with a controllable delay with respect to the XUV pulse. The IR pulse promotes non-sequential transitions paths to the $N=3$ ionization channels, when the pump and probe pulses overlap, as well as sequential transition paths that have as intermediates the many DESs in that energy region. Thanks to the presence of several interfering multi-photon ionization-excitation paths, a coherence between degenerate opposite-parity ionic states does now emerge. 

The XUV pump pulse employed in the simulation has a Gaussian temporal profile, with central frequency $\hbar\omega_\XUV=72.0$~eV (2.646~a.u.), a duration of 970~as (full width at half maximum of the envelope of the intensity, fwhm$_\XUV$), and a peak intensity $I_\XUV$=0.1~TW/cm$^2$. The IR probe pulse has a cosine-squared temporal profile, with central frequency $\hbar\omega_\IR=1.55$~eV (0.057~a.u.), an entire duration of 10.66~fs (fwhm$_\IR\approx$3.77~fs), and peak intensity $I_\IR=$1~GW/cm$^2$.

For case 1 (no $N=4$ channels) the electronic configuration basis comprises, beyond the minimal set of close-coupling channels $N\ell\epsilon_{\ell'}$ with $N\leq 3$, also the full-CI set of configuration $n\ell n'\ell'$ constructed from all the localized orbitals with orbital angular momentum $\ell,\ell'\leq 5$, and total angular momentum $L$ up to 3. The overall size of the {$^1$L$^\pi$} spaces, with $L=0$, $1$, $2$ and $3$, are 4766, 7147, 7946, 7948, respectively, for a total size of 27807. The energy of the ground state is $E_{\mathrm{g}}=-2.8866\,308$~a.u. 

For case 2, the basis includes also the $N=4$ close-coupling channels, which brings the size of the {$^1$L$^\pi$} spaces with $L=0$, $1$, $2$ and $3$, to 7940, 12700, 15090, 15093, respectively (total size 50823). The energy of the ground state changes only marginally, $E_{\mathrm{g}}=-2.8873\,340$~a.u. 

\section{Results}\label{sec:Results}

\begin{figure}[hbtp!]
\begin{center}
\includegraphics[width=\columnwidth]{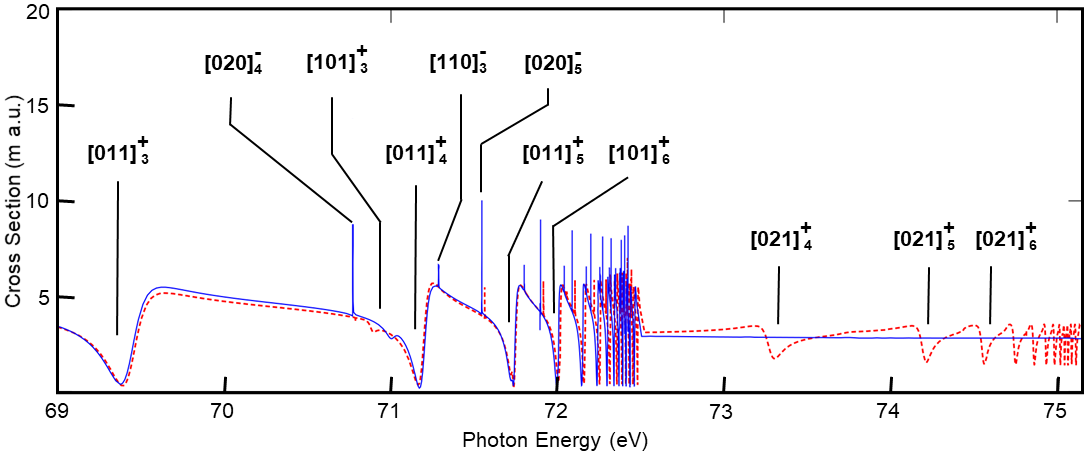}
\caption{\label{fig:Resonances}
Comparison of the energy-resolved total photoionization cross section of the helium atom between the $N=2$ and $N=4$ threshold, computed with (dashed red line) and without (blue line) the $N=4$ close-coupling channels.}
\end{center}
\end{figure}

In this section we analyze the effect of above-threshold resonances on the loss of ionic coherence between opposite parity states. We compare the two cases, with and without the inclusion of $(N+1)_Ln_\ell$ resonances, with focus on the ionization channels $N_L^{'} \epsilon_\ell^{'}$ above $N'=3$. The XUV is broad enough to coherently populate multiple resonances below and above $N=3$ ionization threshold. The central energy of the XUV pulse is chosen not to populate any state below the $N=2$ threshold, where resonances have a large cross-section. The one-photon photoionization amplitudes for the two basis are compared in Figure~\ref{fig:Resonances}; resonances above $N=3$ are present in one case (red dashed line) and absent in the other (blue line). 
The parameters of the resonances below the $N=3$ case change only slightly between the two cases.

In both cases, the XUV pulse populates the DESs below the N=3 threshold, which the IR probe pulse can probe before their Auger decay. The combined absorption of an XUV photon and the additional exchange of IR photons creates a metastable wavepacket that evolves in time with the beating frequencies $\omega_{ij}=E_i-E_j$ between the multiple autoionizing states coherently populated by the pulse sequence. When the atom is ionized, the signs of these beatings can be found in the time delay scan of the photoelectron distribution. In case 2, where we populate resonances above as well as below the $N=3$ threshold, we observe additional beating frequencies, between the $N=4$ resonances and between the $N=3$ resonances with those above the threshold.
\begin{figure*}[hbtp!]
\includegraphics[width=0.9\textwidth]{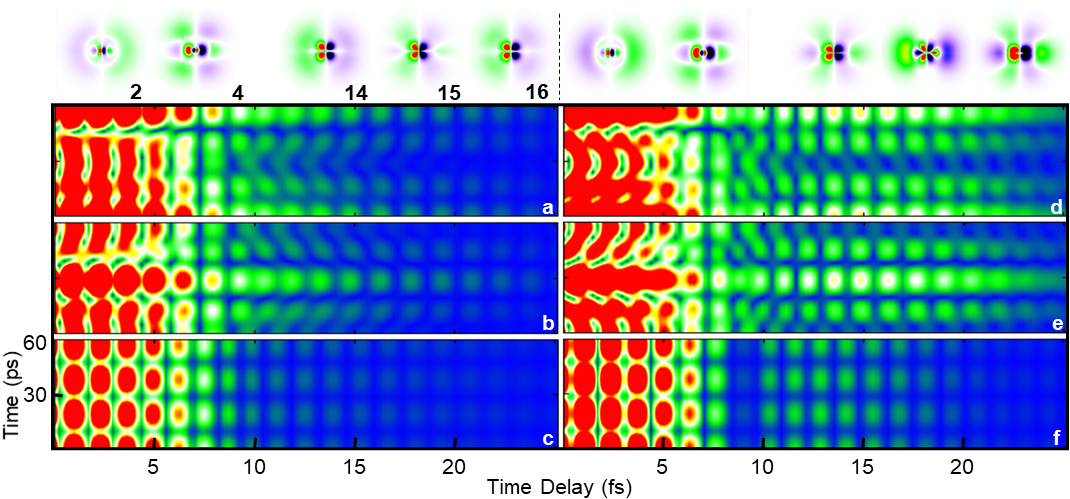}
\caption{\label{fig:PartialCoherence} Partial coherence between off-diagonal terms contributing to dipole is compared for the two cases. left panel is the case 1, whereas right panel displays the same quantities for the case 2. Along the time delay axis, the modulation is different for two cases, whereas, along the real time, the periodicity is similar in two panels. However, periodicity of each term can be different as splitting that contribute to periodicity is different for each coherence pair. Asymmetry between the charge distribution is compared in the top line. The first two snaps at 2 and 4~fs belong to region where pulses overlap, whereas the other three at 14, 15 and 16~fs belongs to a delay where Ir-probe arrives after the XUV-pump. The distribution is very different when pulses don't overlap due to the presence of the additional resonances.}
\end{figure*}

A combination of attosecond XUV-pump and an IR-probe pulses cause the shake-up ionization of helium through several multiphoton paths, some of which involve intermediate resonances. The interference between direct and multi-photon ionization paths gives rise to a partial coherence between states with opposite parity. The ensemble of He$^+$ ions with principal quantum number $N=3$ retains coherence between the $3s$ and $3p$~$(m =0)$ states, the $3p$ and $3d$~$ (m = 0, 1)$ states, and between the $3s$ and $3d$ states. The latter coherence, however, does not manifest itself in the expectation value of the ionic dipole. The framed panels in Figure~\ref{fig:PartialCoherence} show the coherence terms $g_{3p_1,3d_1}(\tau,t)$ (panels a, d), $g_{3p_0,3d_0}$ (panels b, e), $g_{3s,3p_0}(\tau,t)$ (panels c, f), plotted as a function of both the pump-probe delay $\tau$ (on a femtosecond timescale) and the real time $t$ (on a picosecond time scale), computed in case 1 ($N=4$ channels excluded, panels a, b, c) and in case 2 ($N=4$ channels included, panels d, e, f). 

At the low intensity of the IR considered here ($I_{\textsc{IR}}=10^9$\,W/cm$^2$), the dominant cause for the coherence between even and odd states is the overlap of the direct one-photon (XUV) amplitude and the two-photon (XUV+IR) ATI amplitudes above the $N=3$ threshold. It is natural, therefore, that all the coherences are modulated, as a function of the pump-probe delay, at twice the frequency of the IR, throughout the time delay scan. When the two pulses overlap, the non-sequential path can contribute to the two-photon transitions, which justifies the stronger amplitude of the modulation. When the two pulses do not overlap, on the other hand, the absorption of an IR photon can only take place from one of the autoionizing states either above or below the threshold, and hence the coherence exhibits weaker modulations. In fact, the quite stronger contrast of the coherence in the sequential regime in case 2 compared to case 1 suggests either that the two photon amplitude mediated by the $[021]_4^+$ state is stronger than the one mediated by $[011]_4^+$, or that the ATI transitions mediated by the $[011]_n^+$ doubly excited states are much stronger in the presence, in the final state of $N=4$ $S$ and $D$ resonances~\cite{Burgers1995}. The second most prominent change observed when the $N=4$ channels are included is the appearance of a prominent bump in the coherence for time delays between 10~fs and 20~fs, and a clear dip in the coherence around $\tau=9$~fs. We attribute this effect to the multiphoton transitions mediated by the $[021]^+_4$ and $[021]^+_5$ {$^1$P$^o$} doubly excited states, whose lifetimes of 6.7~fs and 11.2~fs, respectively~\cite{Rost_1997}, compares well with the characteristic time-delay duration of the extended signal, which supports the interpretation that it is indeed the intermediate $N=4$ resonance with $L=1$ to enhance coherence, rather than the final resonances with $L=0,\,2$. The minimum at 9~fs is arguably due to a destructive interference between the polarization effect of the ion induced by the tail of the IR and the resonant multiphoton transition. 

The top row in Fig.~\ref{fig:PartialCoherence} shows the asymmetry of the ionic charge density $\rho(x,0,z;\tau)-\rho(x,0,-z;\tau)$, immediately after the end of the pulses, as a function of the time delay, where $\rho(x,y,z;\tau)$ is the charge density, and the external fields are polarized along $z$. In the figure, the $x$ and $z$ axes point up and right, respectively. This quantity illustrates other two aspects of the control of the ion coherence. When the two pulses overlap, the inclusion of the $N=4$ resonances does not have any major effect. This circumstance confirms that, up to $\sim$5~fs, the polarizing effect of the IR probe pulse dominates. At higher values of the delay, in absence of the $N=4$ resonances the density asymmetry experiences only minor changes. When the $N=4$ resonances are included, at the peak of the coherence revival, the density asymmetry changes drastically, inverting its polarization twice between $\tau=$14, 15, and 16~fs.

\begin{figure}[hbtp!]
\includegraphics[width=0.9\columnwidth]{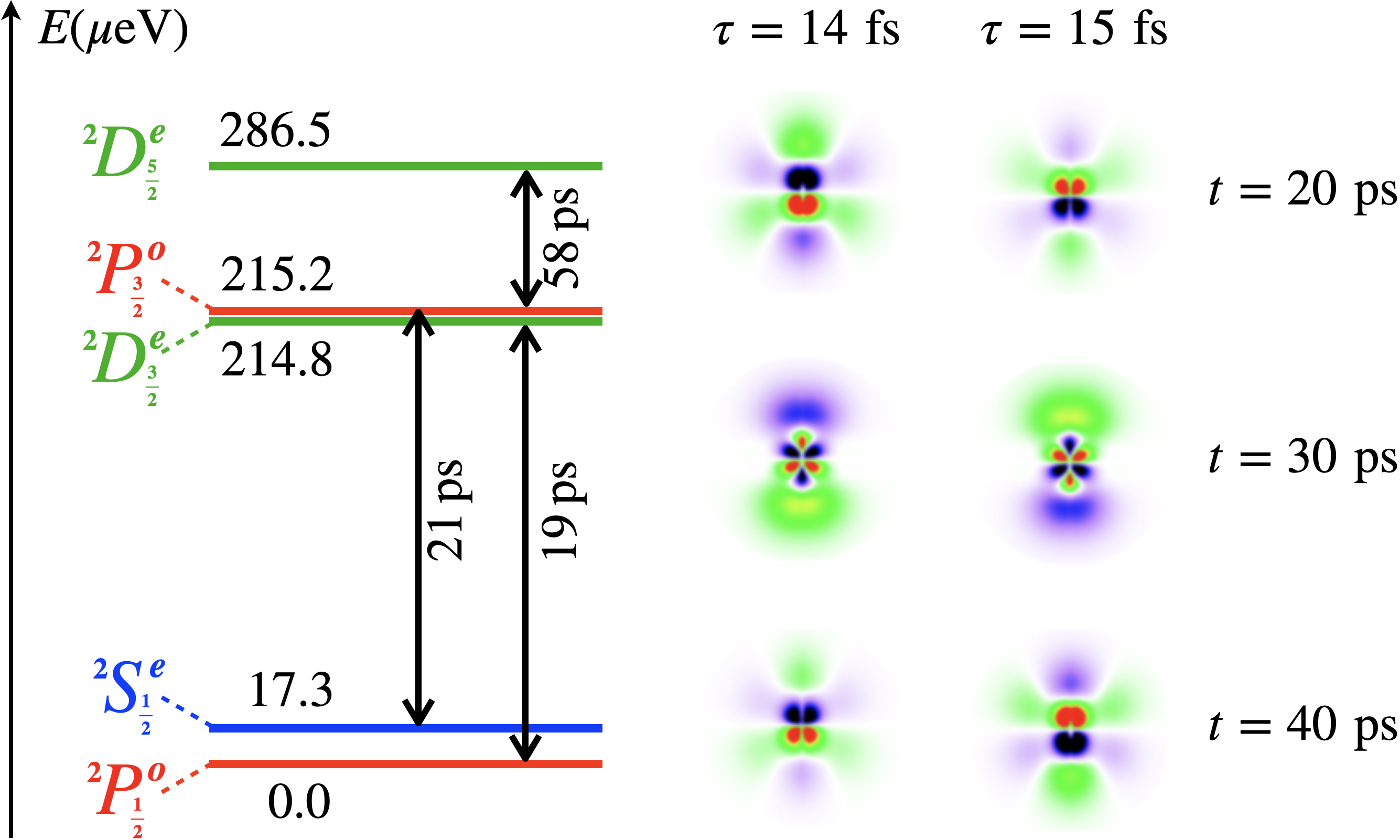}
\caption{\label{fig:EnergySchemeN=3} Energy diagram of the five levels of the He$^+$ states, with principal quantum number $N=3$. The values are computed from the energy levels published by NIST in CODATA~\cite{CODdataNIST}. The number next to each level indicate its relative energy in $\mu$~eV, with respect to the lowest level, $3p\,({^2P^o_{\frac{1}{2}}})$. The fine splitting between these level, which are excited in a partially coherence way by the pump-probe ionization process, causes the charge distribution of the ion, and its electric dipole moment, to fluctuate in real time, on a picosecond timescale. The times next to the vertical arrows are the period of the three fastest oscillations between opposite-parity states, which are observable in the time evolution of the dipole moment of the ionic ensemble. The figures on the right illustrate, for two different time delays, $\tau=14$~fs and $\tau=15$~fs, when the $N=4$ channels are included (case 2, see text for details), the real time evolution of the charge distribution in the $xz$ plane, on a picosecond timescale.}
\end{figure}

As discussed in Sec.~\ref{sec:TheoreticalMethods}, when the fine structure is taken into account, the $N=3$ states of the He$^+$ ion are partly resolved in energy, and hence the charge distribution of the subset of the ions with $N=3$ is no longer stationary (cmp Eq.~\eqref{eq:TDDM}). Figure~\ref{fig:EnergySchemeN=3} illustrates the relative position of the $N=3$ sublevels, and the characteristic picosecond duration of the beatings they induce. Samples of the charge density computed in case 2, in the case of non-overlapping pulses, show how the polarization can change substantially as a result of the angular momentum precession.
The $^2S^e_{\frac{1}{2}}-^2P^o_{\frac{3}{2}}$ and $^2P^o_{\frac{1}{2}}-^2D^e_{\frac{3}{2}}$ splittings are responsible for a periodicity of about $20$~ps in the dipole signal, whereas the $^2P^o_{\frac{3}{2}}-^2D^e_{\frac{5}{2}}$ splitting causes a longer beating, with period $\sim 60$~ps. 

While coherence is not directly measurable, the dipole moment, closely related to coherence between states of opposite parity, is. 
\begin{figure}[hbtp]
\begin{center}
\includegraphics[width=0.9\columnwidth]{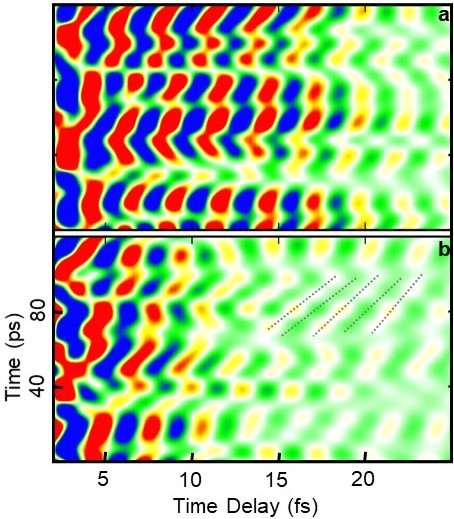}
\caption{\label{fig:dipole}
The average value of the dipole moment is plotted as a function of time delay and real time for case 2(top panel) and case 1 (bottom panel). The effect of the IR is more pronounced in the top panel with larger spacial distribution of the metastable wavepacket due to inclusion of added resonances. The periodicity in time is similar with no single exact periodicity due to multiple frequencies. }
\end{center}
\end{figure}
Figure~\ref{fig:dipole} shows the ionic dipole as a function of pump-probe delay and real time, in both case 1 (panel b) and case 2 (panel a). As commented above, the dipole fluctuates as a function of real time with two dominant frequency components, with period of $\simeq20$~fs and $\simeq 60$~fs. When the pump and probe pulses overlap, the $\simeq 60$~ps period dominates. A similar phenomenon was observed in the dipole of He$^+$ in the $N=2$ level, in pump-probe simulations with a much more intense IR pulse, where a clear checkerboard structure appeared in the dipole spectrum, with the time and time-delay dependence being essentially independent~\cite{Mehmood2021}. Here, this simple checkerboard structure is much less pronounced, due to the comprative weakness of the IR field. 

In the region of positive time delays when the pump and probe pulses do not overlap ($\tau>7$~fs), the modulation of the signal reflects the beating between direct and resonant ionization amplitudes, which are broadly modulated by the $\omega$ frequency, corresponding to the absorption of one IR photon, as well as the beating between two-photon amplitudes mediated by different resonances, which can interfere owing to the finite energy bandwidth of the IR pulse.
The relative phase of the DESs is encoded in the ion's permanent dipole moment, at the end of the pulse sequence, which oscillates as a function of the pump-probe delay on a femtosecond time scale. The dipole also oscillates as a function of real time, on a picosecond timescale. When the pulses do not overlap, in case 1 (panel b), when the $N=4$ channels are not included, we observe a interesting phenomenon: in certain regions of the time/time-delay domain, e.g., $\tau\in[15,22]$~fs, $t\in[60,90]$~ps, the phase of the femtosecond beating maps linearly to that of the picosecond beating, with the period being amplified by a factor of about $6000$. Such magnification, if observable, would allow one to determine the polarization of the ionic ensemble at its inception, with sub-femtosecond precision, from a microwave spectroscopy measurement conducted with picosecond resolution. As panel a shows, the more realistic case in which the $N=4$ channels are included is quite less regular. While the phase of the femtosecond and picosecond beating are not independent, the regions in which they exhibit an approximately linear dependence are much smaller.

\begin{figure}[hbtp]
\begin{center}
\includegraphics[width=0.9\columnwidth]{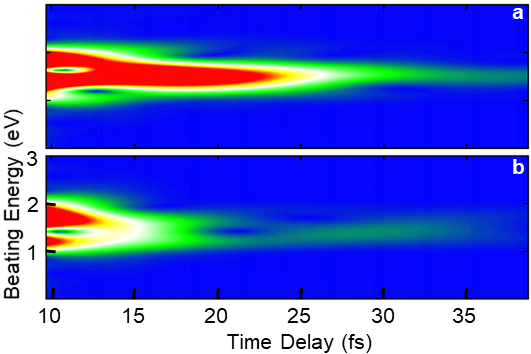}
\caption{\label{fig:FourierTimeDelay}
Dipole moment is comprised of the sum of coherence between opposite states. Its Fourier transform along the delay axis exhibit the beating frequencies of different pairs of DES. When the resonances that lie above the $N=3$ threshold are included, we observe an additional peak at energy of ~2eV which correspond to a beating between a resonance from below and above the threshold.}
\end{center}
\end{figure}

Figure~\ref{fig:FourierTimeDelay} shows the window Fourier transform of the dipole moment with respect to the time delay, 
\begin{equation}
\tilde{\mu}(\tau_w,\omega_\tau) = \frac{1}{\sqrt{8\pi^3}\sigma_w}\int d\tau\,e^{i\omega_\tau \tau-(\tau-\tau_w)^2/2\sigma^2_w} \mu(\tau),
\end{equation} 
where $\sigma_w=1.8$~fs. All the $^1P^o$ DESs below the N=3 that are common to the two cases, giving rise to beatings with frequencies between 1 and 2~eV. The inclusion of the $N=4$ channel results in a much stronger signal between 15 and 30~fs, at the central IR energy ($\simeq 1.55$~eV), which supports the interpretation that, at large delays, the dipolar coherence comes predominantly from the sequential two-photon transitions (one XUV plus one IR) mediated by the $N=4$ resonances, and only to a smaller extent by those mediated by the $N=3$ resonances.
      

\section{Reconstruction of Ionic Coherence}\label{sec:Reconstruction}
As described in the previous section, the spin-orbit interaction induces slow fluctuations into the stationary dipole. Each one of the five beating frequency $\Omega_i$ in the dipole of the $N=3$ ionic ensemble originates from the energy difference between fine-structure states with opposite parity. The dipole can be expressed as a real function of these frequencies, 
 \begin{equation}\label{GEN_DMZ}
  \mu(\tau,t) \propto \sum_{i=1}^{5} c_i(\tau) e^{i\Omega_i t} + c.c.,
\end{equation}
with complex amplitudes $c_{i}$ that depend on the pump-probe delay. These parameters can be extracted from the beating measured in any sufficiently long finite interval, such as the signal recorded in a time-resolved microwave-spectroscopy measurement of the dipole fluctuation. To illustrate the procedure, let's consider a positive square-integrable window function, $W(t)$, that is vanishingly small at any time prior to the end of the pump-probe sequence, such as $W(t)=\phi[(t-t_{W1})/\sigma_W]\phi[(t_{W2}-t)/\sigma_W]$, where $\phi(x) = \frac{1}{\sqrt{2\pi}}\int_{-\infty}^x e^{-t^2/2}dt$ is the normal cumulative distribution function. The associated Window Fourier Transform (WFT) is $\bar{f}(\omega)=\int dt f(t) W(t) e^{-i\omega t}$. The WFT of Eq.~\eqref{GEN_DMZ}, therefore, becomes
\begin{dmath}\label{WFT2}
  \bar{\mu}(\tau,\omega) \propto \sum_{i=1}^{5} [c_{i}(\tau) \widetilde{W}(\Omega_{j}-\omega) +  c_{i}^*(\tau) \widetilde{W}(\Omega_{j}+\omega)],
\end{dmath}
where $\bar{f}(\omega)=\int dt f(t) e^{-i\omega t}$ is the ordinary FT, and the first and the second term corresponds to positive and negative frequency respectively. If the duration of the window step, $\sigma_W$, and plateau, $t_{W2}-t_{W1}$, are much larger than the beating period of any two frequencies $\Omega_i$ and $\Omega_j$, then Eq.~\eqref{WFT2} exhibits an isolated peak for each frequency $\omega=\Omega_j$. Indeed, the FT of the $W(t)$ itself is strongly localized at $\omega=0$. In these conditions, 
\begin{equation}\label{WFT3}
  \bar{\mu}(\tau,\Omega_i) \propto c_{i}(\tau)\widetilde{W}(0),\qquad c_{i}(\tau)= \frac{\bar{\mu}(\tau,\Omega_i)}{\bar{\mu}(\tau,\Omega_j)}c_j(\tau),
\end{equation}
which means that we can reconstruct the amplitude and phase of each oscillation frequency, relative to one of them, used as a reference.
In the case of the He$^+$ $N=2$ states, this conditions can be easily met, since all the fundamental fine-structure frequencies that are visible in the dipole beatings are well separated from each other. 

If the aforementioned conditions are not met, such as in the case of the $N=3$ He$^+$ states, in which at least two pairs of frequencies are very close to each other, a different algorithm is needed. Let's define $\mathbb{M}_{ij} =\widetilde{W}(\Omega_{i}-\Omega_{j})$, $\mathbb{N}_{ij} =\widetilde{W}(\Omega_{i}+\Omega_{j})$, and $\mathbb{b}_i=\bar{\mu}(\tau,\Omega_i)$, with which Eq.~\eqref{WFT2} can be cast in matrix form, 
\begin{equation}\label{WFT_MAT}
   \mathbb{b} = \mathbb{M} \mathbb{c} + \mathbb{N} \mathbb{c^{\star}}, 
\end{equation}
which can be readily solved by considering separately the real and imaginary part of each vector and matrix, $\mathbb{A}^{\Re}=\Re e\mathbb{A}$, $\mathbb{A}^{\Im}=\Im e\mathbb{A}$,
              \begin{equation}\label{MATRIX_EQ}
                  \left[\begin{array}{c}{\mathbb{c}^{\Re}} \\ {\mathbb{c}^{\Im}}\end{array}\right]=2\left[\begin{array}{cc}{(\mathbb{M}+\mathbb{N})^{\Re}} & {(\mathbb{N}-\mathbb{M})^{\Im}} \\ {(\mathbb{M}+\mathbb{N})^{\Im}} & {(\mathbb{M}-\mathbb{N})^{\Re}}\end{array}\right]^{-1}\left[\begin{array}{c}{\mathbb{b}^{\Re}} \\ {\mathbb{b}^{\Im}}\end{array}\right].
              \end{equation}
This latter equation gives the correct solution even if the different peaks of $\tilde{\mu}(\tau,\omega)$ overlap.

Once the amplitude $|c_i(\tau)|$ and phase $\arg c_i(\tau)$ are known, we must determine the set of density matrices that are compatible with these observations. For extremely simple systems, such as He$^+$ $N=2$ states, the beating coefficients are enough to reconstruct the phase of the density matrix off-diagonal elements. In more complex system, such as the one at hand here, the reconstruction cannot be complete because there are more off-diagonal elements than dipole beating modes. 

The slow fluctuations caused by the relativistic interactions can be accurately modeled taking into account that, in the fine-structure basis for the one-electron He$^+$ system, the energy is diagonal. We call $LS$ the one-electron basis $|L,S,M,\Sigma\rangle$, in which the orbital and intrinsic angular momenta are uncoupled, and $fs$ the basis of the fine-structure states, $|\boldsymbol{\Phi}\rangle = |\mathbf{LS}\rangle\mathbf{U}$, $H|\boldsymbol{\Phi}\rangle=|\boldsymbol{\Phi}\rangle\mathbf{E}$, where $E_{ij}=E_i\delta_{ij}$ is a diagonal matrix with the energy of the fine-structure states on the diagonal, and $\mathbf{U}$ is a unitary matrix whose elements are Clebsh-Gordan coefficients.
The density matrix, in the $LS$ basis, then, has the simple time dependence
\begin{equation}\label{UNITRAY_LS_J}
\boldsymbol{\rho}^{LS}(\tau, t) = \mathbf{U} e^{- i \mathbf{E}t} \boldsymbol{\rho}^{fs} (\tau, 0) e^{ i  \mathbf{E}t} \mathbf{U}^{\dagger},
\end{equation}
where $\boldsymbol{\rho}^{fs}(\tau,0)$ is the representation of the density matrix in the fine-structure basis.
As a consequence, the dipole moment can be written as
    \begin{equation}\label{UNITRAY_LS_J}
      \mu(\tau, t) = \mathrm{tr} \left[\mathbf{U} e^{- i \mathbf{E}t} \boldsymbol{\rho}^{fs} (\tau, 0) e^{ i  \mathbf{E}t} \mathbf{U}^{\dagger} \boldsymbol{\mu}^{LS}\right],
    \end{equation}
or, using the cyclic property of the trace, as
\begin{equation}\label{UNITARY_back}
    \mu(\tau, t) = \mathrm{tr}[\boldsymbol{\rho}^{LS} (\tau, 0) \mathbf{U} e^{ i  \mathbf{E}t}\mathbf{U}^{\dagger}\boldsymbol{\mu}^{LS} \mathbf{U} e^{- i  \mathbf{E}t} \mathbf{U}^{\dagger} ]
\end{equation}
   
\begin{figure}[hbtp!]
\begin{center}
\includegraphics[width=0.9\columnwidth]{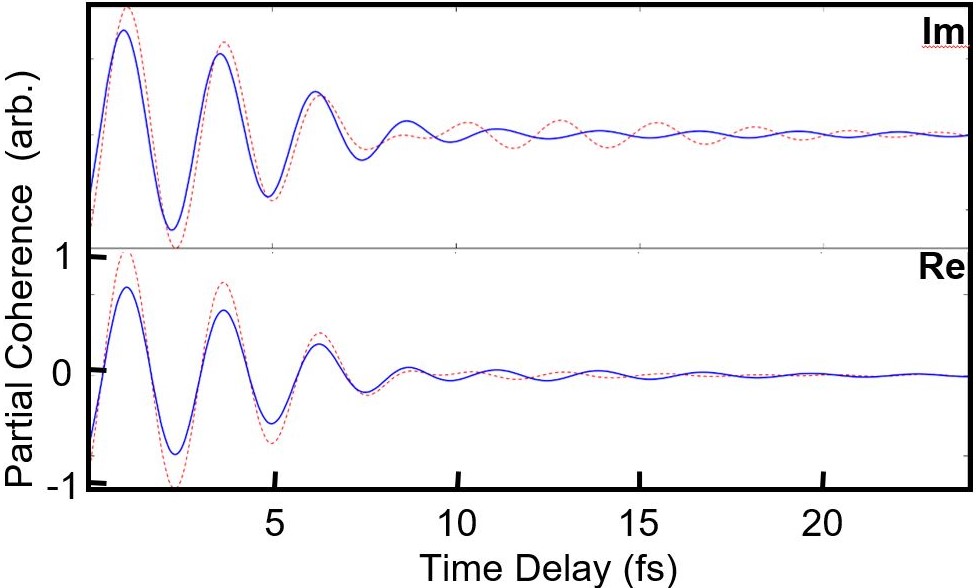}
\caption{\label{fig:Rho3s3p}
Reconstructed real and imaginary parts of the coherence term between $3s$ and $3p_0$ excluding (continuous blue line) and including (dashed red line) the $N=4$ channels. The real and imaginary part are almost in phase.}
\end{center}
\end{figure}

\begin{figure*}[hbtp!]
\begin{center}
\begin{tabular}{lcr}
\includegraphics[width=0.8\columnwidth]{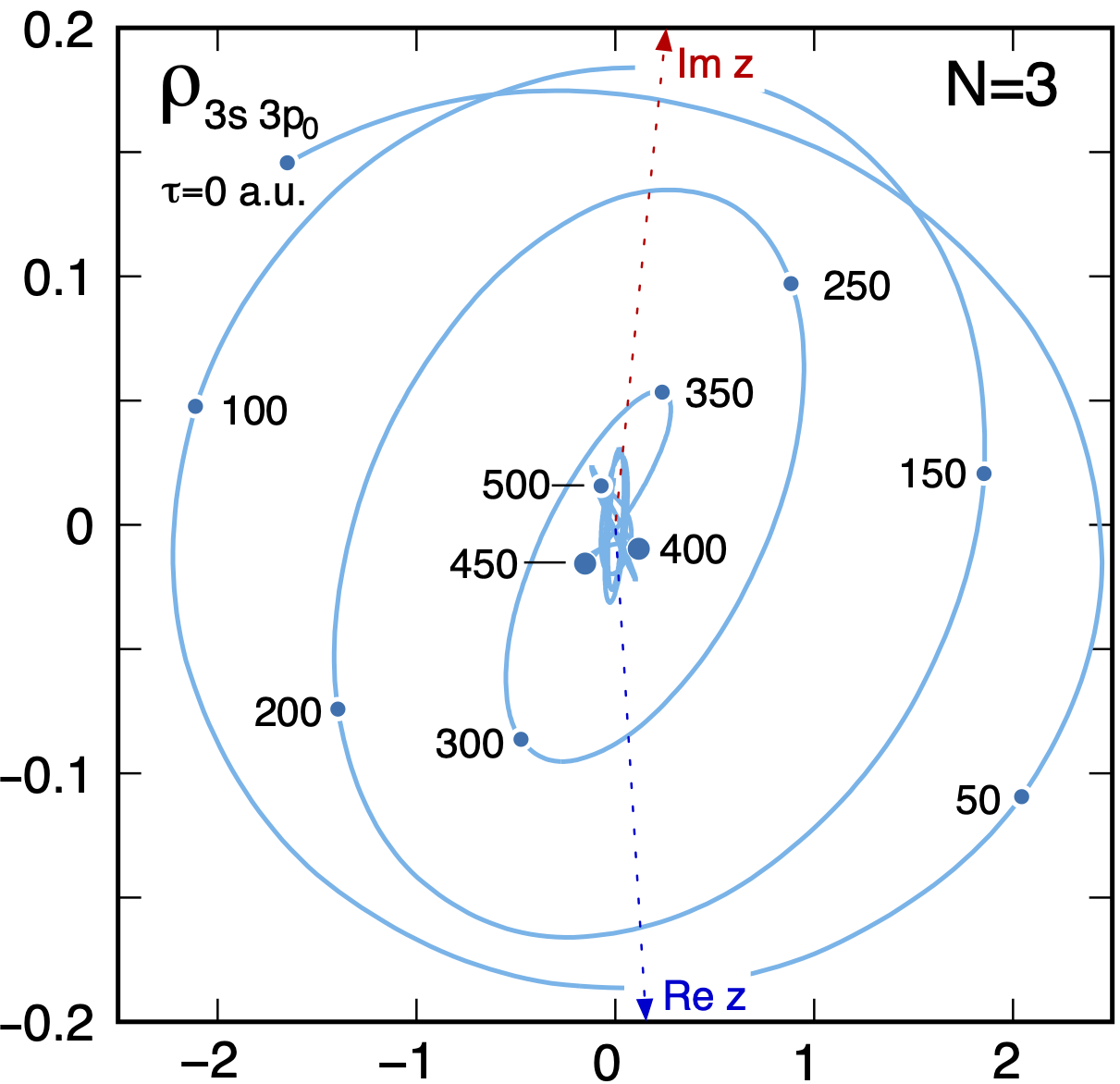}&\phantom{space}&
\includegraphics[width=0.8\columnwidth]{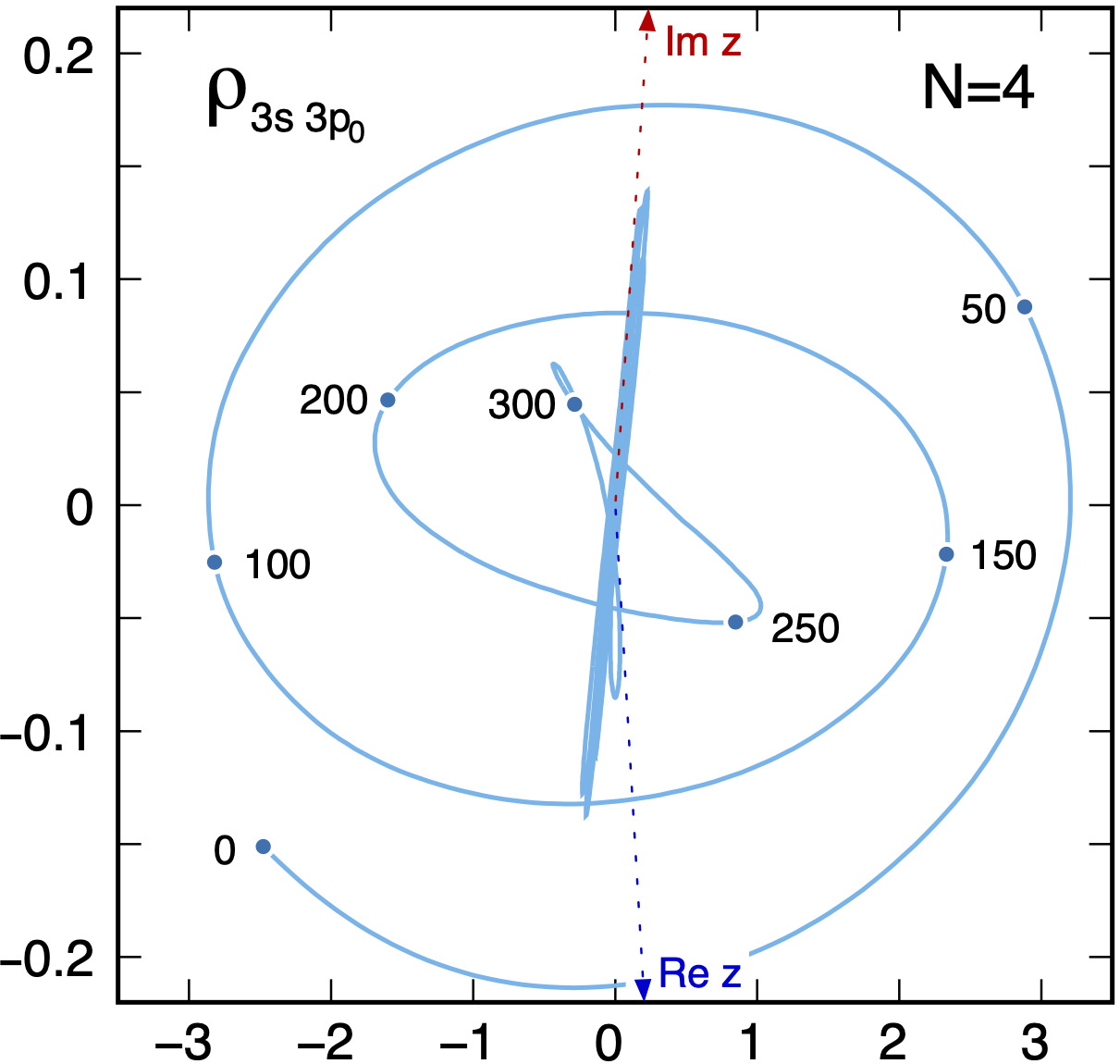}
\end{tabular}
\caption{\label{fig:Rho3s3p_Polar}
Polar representation of the trajectory of $e^{i\phi}\,\rho_{3s3p_0}(\tau)$, as a function of the time delay $\tau$. The two axes are in units of $10^{-3}$. In the $N=3$ case, $\phi=-52^\circ$, whereas for $N=4$ $\phi=-47^\circ$. Notice that the vertical and horizontal axes are on different scale. The dashed line indicate the orientation of the original real and imaginary axes in this representation. When the two pulses overlap, $\rho_{3s3p_0}$ has opposite helicity in the $N=3$ and $N=4$ cases.}
\end{center}
\end{figure*}

\begin{figure}[hbtp]
\begin{center}
\includegraphics[width=\columnwidth]{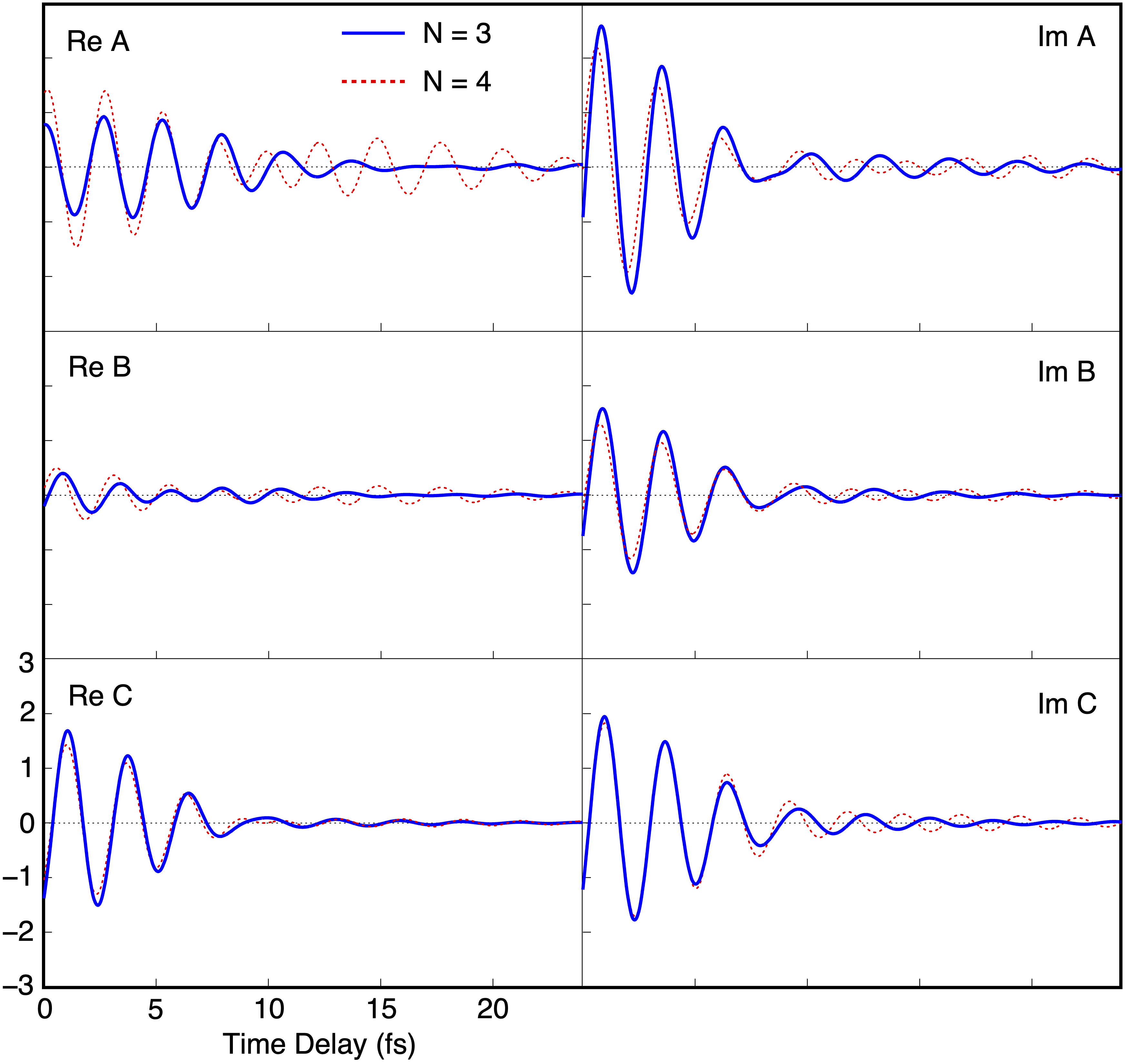}
\caption{\label{fig:ParticularSolutionNew}
Real and Imaginary part of Eqns.~\eqref{eq:Reconstruct3}-\eqref{eq:Reconstruct5} are plotted are plotted as a function of time delay between the pump and the probe pulses.}
\end{center}
\end{figure}

Let's call $\mathbf{P}_a$ the projector on the fine-structure state $a$ in the LS basis, $[\mathbf{P}_a]_{ij}=\mathbf{U}_{ia}\mathbf{U}_{ja}^*$, and let's introduce the matrix $\boldsymbol{\Omega}^{ab} =\mathbf{P}_a\boldsymbol{\mu}^{LS}\mathbf{P}_b$, where $a$ and $b$ are indices of spin-coupled states. Equation~\eqref{UNITARY_back} can be expressed as, 
\begin{equation}\label{eq:Dipole_concise_form}
      \mu(\tau, t) = \sum_{ab}  \mathrm{tr}[\boldsymbol{\rho}^{LS}(\tau,0)\, \boldsymbol{\Omega}^{ab}]\,\,e^{ i  \omega_{ab}t},
\end{equation}
where $\omega_{ab}=E_a-E_b$.
Following the procedure describe above from Equation~\eqref{GEN_DMZ} through~\eqref{MATRIX_EQ}, we can extract the complex coefficients $z_{ab}$ of terms in \eqref{eq:Dipole_concise_form},
\begin{equation}\label{Dipole_projection}
\mathrm{tr}[ \boldsymbol{\rho}^{LS}(\tau,0)\, \boldsymbol{\Omega}^{ab}] = z_{ab}(\tau).
\end{equation}
We can regard \eqref{Dipole_projection} as a system of linear equations for the unknowns $\boldsymbol{\rho}^{LS}(\tau,0)$, where $\boldsymbol{\Omega}^{ab}$ is known analytically and $z_{ab}$ is measured experimentally. Since the dipole moment is sensitive only to coherences between states with opposite parity, it is clear that some of the matrix elements of $\boldsymbol{\rho}$ are not constrained by \eqref{Dipole_projection}. In particular, none of the diagonal terms are. To solve~\eqref{Dipole_projection} systematically, let's express the density matrix as
\begin{equation}\label{eq:DensitySplit}
\begin{split}
\boldsymbol{\rho}&=\sum_i\mathbb{I}^{ii}\rho_{ii}+\sum_{i<j}\left[(\mathbb{I}^{ij}+\mathbb{I}^{ji})\rho_{ij}^\Re + i (\mathbb{I}^{ij}-\mathbb{I}^{ji})\rho_{ij}^\Im\right]
\end{split}
\end{equation}
where $(\mathbb{I}^{ij})_{kl}=\delta_{ik}\delta_{jl}$, $\rho_{ij}^\Re =\Re e(\rho_{ij})$, and $\rho_{ij}^\Im =\Im m(\rho_{ij})$, and the indices $i$ and $j$ correspond the LS states, in which spin and orbital angular momenta are not coupled.
By replacing~\eqref{eq:DensitySplit} into~\eqref{eq:Dipole_concise_form}, we obtain
\begin{equation}\label{Trace_free_proj}
\begin{split}
z_{ab} &=\phantom{i} {\sum_{i<j}}'\mathrm{tr}[(\mathbb{I}^{ij} + \mathbb{I}^{ji}) \boldsymbol{\Omega}^{ab}] \, \rho_{ij}^{\Re}+\\  
&+i {\sum_{i< j}}'  \mathrm{tr}[(\mathbb{I}^{ij} - \mathbb{I}^{ji}) \boldsymbol{\Omega}^{ab}] \rho_{ij}^{\Im},
\end{split}
\end{equation}
where we used the fact that $\mathrm{tr}[\mathbb{I}^{ii}\boldsymbol{\Omega}^{ab}]=0$, and the prime in the summation indicates that we skip over pairs of states $i$ and $j$ that have the same parity. To avoid redundancy, we can assume that $a$ is an even state and $b$ is odd (only states with opposite parity beat with each other). It is also convenient to order the LS basis such that the even states precede the odd ones. If $i$ and $j$ have even and odd parity, respectively, then the trace $\mathrm{tr}[\mathbb{I}^{ji}\boldsymbol{\Omega}^{ab}]=0$. Therefore, we can rewrite~\eqref{Trace_free_proj} as
\begin{equation}\label{Reduced_reconstruction}
\begin{split}
z_{ab} &=\phantom{i} \sum_i^e\sum_{j}^o\mathrm{tr}[\mathbb{I}^{ij} \boldsymbol{\Omega}^{ab}] \, \rho_{ij}
\end{split}
\end{equation}
The relation has now been cast in the form of a linear system for the complex upper-diagonal components of the density matrix between opposite-parity states.
Let's define the superindexes $I=(i,j)$ and $A=(a,b)$, and introduce the notation $M_{AI}=\mathrm{tr}[\mathbb{I}^{ij} \boldsymbol{\Omega}^{ab}]$ and  $\rho_I = \rho_{ij} (\tau, t=0)$ and $z_A=z_{ab}$. The system~\eqref{Reduced_reconstruction}, then, becomes
\begin{equation}\label{eq:SimpleEquationForRho}
    \mathbf{M}\boldsymbol{\rho}=\mathbf{z}
\end{equation}
The matrix $\mathbf{M}$ is rectangular with 5 rows and 6 columns. Since $\mathbf{M}$ has more columns than rows, the general solution $\boldsymbol{\rho}$ can only be written up to an arbitrary solution of the associated homogeneous system, $\mathbf{A}\boldsymbol{\rho}_h=0$,
\begin{equation}
   \boldsymbol{\rho}=\boldsymbol{\rho}_{p} + \boldsymbol{\rho}_h,
\end{equation}
where $\boldsymbol{\rho}_{p}$ is a particular solution.
To determine the particular solution and the linear space of homogeneous solutions, we can solve the problem
\begin{equation}
 \mathbf{M}^\dagger\mathbf{M}\boldsymbol{\rho}=\mathbf{M}^\dagger\mathbf{z},
\end{equation}
where $\mathbf{S}=\mathbf{M}^\dagger\mathbf{M}$ is a positive definite, symmetric real matrix with rank smaller than its dimension. $\mathbf{S}$ and $\mathbf{M}$ have the same null space, which is spanned by the $N_{\mathcal{K}}$ eigenvectors of $\mathbf{S}$, $\mathbf{U}_{\mathcal{K}}$, with eigenvalue zero. We can search for a particular solution of~\eqref{eq:SimpleEquationForRho} in the range of $\mathbf{S}$, $\mathbf{U}_{\mathcal{R}}$, orthogonal to the null space, $\mathbf{U}_{\mathcal{R}}^\dagger\mathbf{U}_{\mathcal{K}}=0$, which has size $N_{\mathcal{R}}=N-N_{\mathcal{K}}$,
\begin{equation}
    \boldsymbol{\rho}_p = \mathbf{U}_{\mathcal{R}}\mathbf{c}.
\end{equation}
This expression leads to the following equation for the set of coefficients $\mathbf{c}$,
\begin{equation}
\mathbf{U}_R^\dagger\mathbf{M}\mathbf{U}_R\mathbf{c}=\mathbf{U}_R^\dagger\mathbf{z},
\end{equation}
which is readily solved,
\begin{equation}
\mathbf{c}=[\mathbf{U}_R^\dagger\mathbf{M}\mathbf{U}_R]^{-1}\mathbf{U}_R^\dagger\mathbf{z}.
\end{equation}
To summarize, the general solution to~\eqref{eq:SimpleEquationForRho} is
\begin{equation}
   \boldsymbol{\rho}(\boldsymbol{\alpha})=\mathbf{U}_R[\mathbf{U}_R^\dagger\mathbf{M}\mathbf{U}_R]^{-1}\mathbf{U}_R^\dagger\mathbf{z} + \mathbf{U}_K\boldsymbol{\alpha},
\end{equation}
where $\boldsymbol{\alpha}=(\alpha_1,\alpha_2,\ldots,\alpha_{N_{\mathcal{K}}})^t$ is a vector of arbitrary complex numbers, with the same dimension as the null space.
The eigenvectors over which the particular solution is expressed represent the linear combination of density-matrix elements that can be reconstructed from an all-optical measurement. The present excitation scheme has a duration of few tens of femtoseconds, i.e., three orders of magnitude smaller than the spin precession period caused by the fine-structure splitting. As long as the electron spin does not affect the excitation process, therefore, the dipole expectation value at the end of the pulses is dictated only by the coherence between the $3s$, $3p_{0}$ and $3p_{m}$, $3d_{m}$  states, whereas the coherence between the $3s$ and the $3\bar{p}_{1}$, $3p_{m}$ and the $3\bar{d}_{m+1}$, $3d_{m}$ and the $3\bar{p}_{m+1}$ states is zero.  At larger times, the non-stationary character of the $3p_{m}$ and $3d_{m}$ configurations emerges, and the dipole moment is observed to oscillate. 
The density-matrix elements that can be determined using these procedure for the $N=3$ He$^+$ states are
\begin{eqnarray}
&&\rho_{3s_{0},3p_0},\qquad \rho_{3s_0,3\bar{p}_1}\label{eq:Reconstruct1}\\
A &=&\sqrt{2} \rho_{3p_{0},3d_{0}} + \rho_{3p_{0},3\bar{d}_{1}}\label{eq:Reconstruct3}\\
B &=&\sqrt{2} \rho_{3\bar{p}_{1},3\bar{d}_{1}} + \rho_{3\bar{p}_{1},3d_{0}}\\
C &=&\frac{1}{\sqrt{2}} \rho_{3p_{0},3\bar{d}_{1}} + \frac{3\sqrt{3}}{2}\rho_{3\bar{p}_{1},3d_{0}}\label{eq:Reconstruct5},
\end{eqnarray}
where, for the suffixes, we use the notation $3\ell_m$ for spin-up states, and $3\bar{\ell}_m$ for spin-down states.
To check the consistency of this method, we compared the five reconstructed quantities in Eqns.~\eqref{eq:Reconstruct1}-\eqref{eq:Reconstruct5} with the same exact quantities computed in our \emph{ab initio} simulations, finding a perfect agreement. 
Figure~\ref{fig:Rho3s3p} shows the reconstructed real and imaginary part of $\rho_{3s_0,3p_0}$, as a function of the pump-probe delay, computed either excluding or including the $N=4$ channels in the calculation. Some aspects of this plot can be noticed. First, the oscillations in the presence of the $N=4$ channels are more pronounced in both the short and mid time-delay region. Second, while the $N=3$ and $N=4$ oscillations are approximately in phase in the delay region where the pump and probe overlap, this is no longer the case in the mid time-delay region, where the coherence in the $N=4$ case is dominated by contribution from the intermediate above-the-threshold resonances. Third, the real and imaginary part of the coherence are almost in phase, which means that the coherence oscillate back and forth through, or near, the origin.
Figure~\ref{fig:Rho3s3p_Polar} highlights this latter aspect in more detail, by representing the function $\rho_{3s3p_0}$ rotated clockwise by $52^\circ$ and by $47^\circ$, in the $N=3$ and $N=4$ cases, respectively, on a scale that approximately equalizes the vertical and horizontal oscillations in the region of the overlapping pulses. Points at specific time delays are highlighted in the figure. In this representation, it is possible to appreciate a qualitative difference between the two cases for the overlapping pulses: in the case of $N=3$, $\rho_{3s3p_0}$ winds clockwise around the origin, whereas, in the case of $N=4$, it goes around the origin counterclockwise.
As mentioned at the beginning, in our case, the quantity $\rho_{3s_0,3\bar{p}_1}$ is identically zero, since we do not have terms in the Hamiltonian employed to simulate the attosecond pump-probe experiment that account for relativistic terms. As noted in~\cite{Mehmood2021}, this experimental observable is sensitive to the influence of relativistic terms in attosecond ionization processes.
Figure~\ref{fig:ParticularSolutionNew} shows the real and imaginary part of the combinations $A$, $B$, and $C$ of pairs of off-diagonal elements of the density matrix, given in Eqns~\eqref{eq:Reconstruct3}-\eqref{eq:Reconstruct5}, for the two different set of channels.
The main features of these combinations are similar to those already discussed for $\rho_{3s3p_0}$, with one notable difference. In the time-delay region of overlapping pulses, for both $B$ and $C$, the real and imaginary parts are close to be in phase, thus resulting in an oscillation of the absolute value of $|B|$ and $|C|$ themselves. In the case of $A$, on the other hand, the real and imaginary part are nearly in quadrature, which suggests that $|A|$ is conserved. Finally, $C$ is the only combination of density matrix elements that does not feature terms between ionic states with the same spin, and it is also the only term whose peak amplitude is not appreciably increased by the inclusion of the $N=4$ states, particularly in the region where the pulses do not overlap, where both $|A|$ and $|B|$ do increase.

\section{Conclusion}\label{sec:Conclusions}
In this work, we have extended to the $N=3$ excited state our study on the control of the He$^+$ ion generated in attosecond pump-probe spectroscopy. We have assessed the role of the resonances above the $N=3$ threshold, which converge to the subsequent $N=4$ threshold, and ascertain that they significantly extend the range of time delays beyond which the ionic coherence is enhanced. This phenomenon is understood to be due to the sequential XUV + IR two-photon amplitudes to the $N=3$ channels with even parity, mediated by $N=4$ intermediate autoionizing states. We also describe a generalization of the reconstruction protocol of the density-matrix off-diagonal elements from the picosecond fluctuation of the ionic dipole, which can in principle be measured with microwave spectroscopy. While the reconstruction cannot be complete, owing to the presence of coherences that do not manifest themselves in the expectation value of the dipole, it does impose well-defined constraints on the coherences between opposite-parity states. In particular, it is possible to use this reconstruction procedure to measure the  opposite-spin $\rho_{3s,3\bar{p}_1}$ coherence with respect to the same-spin $\rho_{3s,3p_0}$ coherence. A finite result for this measurement would in turn quantify the role of relativistic effects during the attosecond ionization event itself.

\begin{acknowledgments}
This work is supported by NSF grant no. 1607588. EL acknowledges support from Swedish Research Council, Grant No. 2020-03315. Special thanks to UCF ARCC for computing time on STOKES super computer.
\end{acknowledgments}


\begin{thebibliography}{57}%
\makeatletter
\providecommand \@ifxundefined [1]{%
 \@ifx{#1\undefined}
}%
\providecommand \@ifnum [1]{%
 \ifnum #1\expandafter \@firstoftwo
 \else \expandafter \@secondoftwo
 \fi
}%
\providecommand \@ifx [1]{%
 \ifx #1\expandafter \@firstoftwo
 \else \expandafter \@secondoftwo
 \fi
}%
\providecommand \natexlab [1]{#1}%
\providecommand \enquote  [1]{``#1''}%
\providecommand \bibnamefont  [1]{#1}%
\providecommand \bibfnamefont [1]{#1}%
\providecommand \citenamefont [1]{#1}%
\providecommand \href@noop [0]{\@secondoftwo}%
\providecommand \href [0]{\begingroup \@sanitize@url \@href}%
\providecommand \@href[1]{\@@startlink{#1}\@@href}%
\providecommand \@@href[1]{\endgroup#1\@@endlink}%
\providecommand \@sanitize@url [0]{\catcode `\\12\catcode `\$12\catcode
  `\&12\catcode `\#12\catcode `\^12\catcode `\_12\catcode `\%12\relax}%
\providecommand \@@startlink[1]{}%
\providecommand \@@endlink[0]{}%
\providecommand \url  [0]{\begingroup\@sanitize@url \@url }%
\providecommand \@url [1]{\endgroup\@href {#1}{\urlprefix }}%
\providecommand \urlprefix  [0]{URL }%
\providecommand \Eprint [0]{\href }%
\providecommand \doibase [0]{https://doi.org/}%
\providecommand \selectlanguage [0]{\@gobble}%
\providecommand \bibinfo  [0]{\@secondoftwo}%
\providecommand \bibfield  [0]{\@secondoftwo}%
\providecommand \translation [1]{[#1]}%
\providecommand \BibitemOpen [0]{}%
\providecommand \bibitemStop [0]{}%
\providecommand \bibitemNoStop [0]{.\EOS\space}%
\providecommand \EOS [0]{\spacefactor3000\relax}%
\providecommand \BibitemShut  [1]{\csname bibitem#1\endcsname}%
\let\auto@bib@innerbib\@empty
\bibitem [{\citenamefont {Krausz}\ and\ \citenamefont
  {Ivanov}(2009)}]{Krausz2009}%
  \BibitemOpen
  \bibfield  {author} {\bibinfo {author} {\bibfnamefont {F.}~\bibnamefont
  {Krausz}}\ and\ \bibinfo {author} {\bibfnamefont {M.}~\bibnamefont
  {Ivanov}},\ }\bibfield  {title} {\bibinfo {title} {{Attosecond physics}},\
  }\href {https://doi.org/10.1103/RevModPhys.81.163} {\bibfield  {journal}
  {\bibinfo  {journal} {Rev. Mod. Phys.}\ }\textbf {\bibinfo {volume} {81}},\
  \bibinfo {pages} {163} (\bibinfo {year} {2009})}\BibitemShut {NoStop}%
\bibitem [{\citenamefont {Pazourek}\ \emph
  {et~al.}(2015{\natexlab{a}})\citenamefont {Pazourek}, \citenamefont
  {Nagele},\ and\ \citenamefont {Burgd\"orfer}}]{RevModPhys.87.765}%
  \BibitemOpen
  \bibfield  {author} {\bibinfo {author} {\bibfnamefont {R.}~\bibnamefont
  {Pazourek}}, \bibinfo {author} {\bibfnamefont {S.}~\bibnamefont {Nagele}},\
  and\ \bibinfo {author} {\bibfnamefont {J.}~\bibnamefont {Burgd\"orfer}},\
  }\bibfield  {title} {\bibinfo {title} {Attosecond chronoscopy of
  photoemission},\ }\href {https://doi.org/10.1103/RevModPhys.87.765}
  {\bibfield  {journal} {\bibinfo  {journal} {Rev. Mod. Phys.}\ }\textbf
  {\bibinfo {volume} {87}},\ \bibinfo {pages} {765} (\bibinfo {year}
  {2015}{\natexlab{a}})}\BibitemShut {NoStop}%
\bibitem [{\citenamefont {Calegari}\ \emph {et~al.}(2016)\citenamefont
  {Calegari}, \citenamefont {Sansone}, \citenamefont {Stagira}, \citenamefont
  {Vozzi},\ and\ \citenamefont {Nisoli}}]{Calegari_2016}%
  \BibitemOpen
  \bibfield  {author} {\bibinfo {author} {\bibfnamefont {F.}~\bibnamefont
  {Calegari}}, \bibinfo {author} {\bibfnamefont {G.}~\bibnamefont {Sansone}},
  \bibinfo {author} {\bibfnamefont {S.}~\bibnamefont {Stagira}}, \bibinfo
  {author} {\bibfnamefont {C.}~\bibnamefont {Vozzi}},\ and\ \bibinfo {author}
  {\bibfnamefont {M.}~\bibnamefont {Nisoli}},\ }\bibfield  {title} {\bibinfo
  {title} {Advances in attosecond science},\ }\href
  {https://doi.org/10.1088/0953-4075/49/6/062001} {\bibfield  {journal}
  {\bibinfo  {journal} {J. Phys. B: At. Mol. Opt. Phys.}\ }\textbf {\bibinfo
  {volume} {49}},\ \bibinfo {pages} {062001} (\bibinfo {year}
  {2016})}\BibitemShut {NoStop}%
\bibitem [{\citenamefont {L{\'{e}}pine}\ \emph {et~al.}(2013)\citenamefont
  {L{\'{e}}pine}, \citenamefont {Sansone},\ and\ \citenamefont
  {Vrakking}}]{Lepine2013}%
  \BibitemOpen
  \bibfield  {author} {\bibinfo {author} {\bibfnamefont {F.}~\bibnamefont
  {L{\'{e}}pine}}, \bibinfo {author} {\bibfnamefont {G.}~\bibnamefont
  {Sansone}},\ and\ \bibinfo {author} {\bibfnamefont {M.~J.~J.}\ \bibnamefont
  {Vrakking}},\ }\bibfield  {title} {\bibinfo {title} {{Molecular applications
  of attosecond laser pulses}},\ }\href
  {https://doi.org/10.1016/j.cplett.2013.05.045} {\bibfield  {journal}
  {\bibinfo  {journal} {Chem. Phys. Lett.}\ }\textbf {\bibinfo {volume}
  {578}},\ \bibinfo {pages} {1} (\bibinfo {year} {2013})}\BibitemShut {NoStop}%
\bibitem [{\citenamefont {L{\'{e}}pine}\ \emph {et~al.}(2014)\citenamefont
  {L{\'{e}}pine}, \citenamefont {Ivanov},\ and\ \citenamefont
  {Vrakking}}]{Lepine2014}%
  \BibitemOpen
  \bibfield  {author} {\bibinfo {author} {\bibfnamefont {F.}~\bibnamefont
  {L{\'{e}}pine}}, \bibinfo {author} {\bibfnamefont {M.~Y.}\ \bibnamefont
  {Ivanov}},\ and\ \bibinfo {author} {\bibfnamefont {M.~J.~J.}\ \bibnamefont
  {Vrakking}},\ }\bibfield  {title} {\bibinfo {title} {{Attosecond molecular
  dynamics: fact or fiction?}},\ }\href
  {https://doi.org/10.1038/nphoton.2014.25} {\bibfield  {journal} {\bibinfo
  {journal} {Nat. Photonics}\ }\textbf {\bibinfo {volume} {8}},\ \bibinfo
  {pages} {195} (\bibinfo {year} {2014})}\BibitemShut {NoStop}%
\bibitem [{\citenamefont {Leone}\ \emph {et~al.}(2014)\citenamefont {Leone},
  \citenamefont {McCurdy}, \citenamefont {Burgd{\"{o}}rfer}, \citenamefont
  {Cederbaum}, \citenamefont {Chang}, \citenamefont {Dudovich}, \citenamefont
  {Feist}, \citenamefont {Greene}, \citenamefont {Ivanov}, \citenamefont
  {Kienberger}, \citenamefont {Keller}, \citenamefont {Kling}, \citenamefont
  {Loh}, \citenamefont {Pfeifer}, \citenamefont {Pfeiffer}, \citenamefont
  {Santra}, \citenamefont {Schafer}, \citenamefont {Stolow}, \citenamefont
  {Thumm},\ and\ \citenamefont {Vrakking}}]{Leone2014}%
  \BibitemOpen
  \bibfield  {author} {\bibinfo {author} {\bibfnamefont {S.~R.}\ \bibnamefont
  {Leone}}, \bibinfo {author} {\bibfnamefont {C.~W.}\ \bibnamefont {McCurdy}},
  \bibinfo {author} {\bibfnamefont {J.}~\bibnamefont {Burgd{\"{o}}rfer}},
  \bibinfo {author} {\bibfnamefont {L.~S.}\ \bibnamefont {Cederbaum}}, \bibinfo
  {author} {\bibfnamefont {Z.}~\bibnamefont {Chang}}, \bibinfo {author}
  {\bibfnamefont {N.}~\bibnamefont {Dudovich}}, \bibinfo {author}
  {\bibfnamefont {J.}~\bibnamefont {Feist}}, \bibinfo {author} {\bibfnamefont
  {C.~H.}\ \bibnamefont {Greene}}, \bibinfo {author} {\bibfnamefont {M.~Y.}\
  \bibnamefont {Ivanov}}, \bibinfo {author} {\bibfnamefont {R.}~\bibnamefont
  {Kienberger}}, \bibinfo {author} {\bibfnamefont {U.}~\bibnamefont {Keller}},
  \bibinfo {author} {\bibfnamefont {M.~F.}\ \bibnamefont {Kling}}, \bibinfo
  {author} {\bibfnamefont {Z.-H.}\ \bibnamefont {Loh}}, \bibinfo {author}
  {\bibfnamefont {T.}~\bibnamefont {Pfeifer}}, \bibinfo {author} {\bibfnamefont
  {A.~N.}\ \bibnamefont {Pfeiffer}}, \bibinfo {author} {\bibfnamefont
  {R.}~\bibnamefont {Santra}}, \bibinfo {author} {\bibfnamefont {K.~J.}\
  \bibnamefont {Schafer}}, \bibinfo {author} {\bibfnamefont {A.}~\bibnamefont
  {Stolow}}, \bibinfo {author} {\bibfnamefont {U.}~\bibnamefont {Thumm}},\ and\
  \bibinfo {author} {\bibfnamefont {M.~J.~J.}\ \bibnamefont {Vrakking}},\
  }\bibfield  {title} {\bibinfo {title} {{What will it take to observe
  processes in 'real time'?}},\ }\href
  {https://doi.org/10.1038/nphoton.2014.48} {\bibfield  {journal} {\bibinfo
  {journal} {Nat. Photonics}\ }\textbf {\bibinfo {volume} {8}},\ \bibinfo
  {pages} {162} (\bibinfo {year} {2014})}\BibitemShut {NoStop}%
\bibitem [{\citenamefont {Leone}\ and\ \citenamefont
  {Neumark}(2016)}]{LeoneNeumark2016}%
  \BibitemOpen
  \bibfield  {author} {\bibinfo {author} {\bibfnamefont {S.~R.}\ \bibnamefont
  {Leone}}\ and\ \bibinfo {author} {\bibfnamefont {D.~M.}\ \bibnamefont
  {Neumark}},\ }\bibfield  {title} {\bibinfo {title} {{Attosecond science in
  atomic, molecular, and condensed matter physics}},\ }\href
  {https://doi.org/10.1039/C6FD00174B} {\bibfield  {journal} {\bibinfo
  {journal} {Faraday Discuss.}\ }\textbf {\bibinfo {volume} {194}},\ \bibinfo
  {pages} {15} (\bibinfo {year} {2016})}\BibitemShut {NoStop}%
\bibitem [{\citenamefont {Nisoli}\ \emph {et~al.}(2017)\citenamefont {Nisoli},
  \citenamefont {Decleva}, \citenamefont {Calegari}, \citenamefont {Palacios},\
  and\ \citenamefont {Mart{\'{i}}n}}]{Nisoli2017}%
  \BibitemOpen
  \bibfield  {author} {\bibinfo {author} {\bibfnamefont {M.}~\bibnamefont
  {Nisoli}}, \bibinfo {author} {\bibfnamefont {P.}~\bibnamefont {Decleva}},
  \bibinfo {author} {\bibfnamefont {F.}~\bibnamefont {Calegari}}, \bibinfo
  {author} {\bibfnamefont {A.}~\bibnamefont {Palacios}},\ and\ \bibinfo
  {author} {\bibfnamefont {F.}~\bibnamefont {Mart{\'{i}}n}},\ }\bibfield
  {title} {\bibinfo {title} {{Attosecond Electron Dynamics in Molecules}},\
  }\href {https://doi.org/10.1021/acs.chemrev.6b00453} {\bibfield  {journal}
  {\bibinfo  {journal} {Chem. Rev.}\ }\textbf {\bibinfo {volume} {117}},\
  \bibinfo {pages} {10760} (\bibinfo {year} {2017})}\BibitemShut {NoStop}%
\bibitem [{\citenamefont {Sansone}\ \emph {et~al.}(2012)\citenamefont
  {Sansone}, \citenamefont {Pfeifer}, \citenamefont {Simeonidis},\ and\
  \citenamefont {Kuleff}}]{Sansone2012}%
  \BibitemOpen
  \bibfield  {author} {\bibinfo {author} {\bibfnamefont {G.}~\bibnamefont
  {Sansone}}, \bibinfo {author} {\bibfnamefont {T.}~\bibnamefont {Pfeifer}},
  \bibinfo {author} {\bibfnamefont {K.}~\bibnamefont {Simeonidis}},\ and\
  \bibinfo {author} {\bibfnamefont {A.~I.}\ \bibnamefont {Kuleff}},\ }\bibfield
   {title} {\bibinfo {title} {{Electron correlation in real time.}},\ }\href
  {https://doi.org/10.1002/cphc.201100528} {\bibfield  {journal} {\bibinfo
  {journal} {Chem. Phys. Chem.}\ }\textbf {\bibinfo {volume} {13}},\ \bibinfo
  {pages} {661} (\bibinfo {year} {2012})}\BibitemShut {NoStop}%
\bibitem [{\citenamefont {Pazourek}\ \emph
  {et~al.}(2015{\natexlab{b}})\citenamefont {Pazourek}, \citenamefont
  {Nagele},\ and\ \citenamefont {Burgd{\"{o}}rfer}}]{PazourekRMP2015}%
  \BibitemOpen
  \bibfield  {author} {\bibinfo {author} {\bibfnamefont {R.}~\bibnamefont
  {Pazourek}}, \bibinfo {author} {\bibfnamefont {S.}~\bibnamefont {Nagele}},\
  and\ \bibinfo {author} {\bibfnamefont {J.}~\bibnamefont {Burgd{\"{o}}rfer}},\
  }\bibfield  {title} {\bibinfo {title} {{Attosecond chronoscopy of
  photoemission}},\ }\href {https://doi.org/10.1103/RevModPhys.87.765}
  {\bibfield  {journal} {\bibinfo  {journal} {Rev. Mod. Phys.}\ }\textbf
  {\bibinfo {volume} {87}},\ \bibinfo {pages} {765} (\bibinfo {year}
  {2015}{\natexlab{b}})}\BibitemShut {NoStop}%
\bibitem [{\citenamefont {Ciappina}\ \emph {et~al.}(2017)\citenamefont
  {Ciappina}, \citenamefont {P{\'{e}}rez-Hern{\'{a}}ndez}, \citenamefont
  {Landsman}, \citenamefont {Okell}, \citenamefont {Zherebtsov}, \citenamefont
  {Förg}, \citenamefont {Schötz}, \citenamefont {Seiffert}, \citenamefont
  {Fennel}, \citenamefont {Shaaran}, \citenamefont {Zimmermann}, \citenamefont
  {Chac{\'{o}}n}, \citenamefont {Guichard}, \citenamefont {Zaïr},
  \citenamefont {Tisch}, \citenamefont {Marangos}, \citenamefont {Witting},
  \citenamefont {Braun}, \citenamefont {Maier}, \citenamefont {Roso},
  \citenamefont {Krüger}, \citenamefont {Hommelhoff}, \citenamefont {Kling},
  \citenamefont {Krausz},\ and\ \citenamefont {Lewenstein}}]{Ciappina_2017}%
  \BibitemOpen
  \bibfield  {author} {\bibinfo {author} {\bibfnamefont {M.~F.}\ \bibnamefont
  {Ciappina}}, \bibinfo {author} {\bibfnamefont {J.~A.}\ \bibnamefont
  {P{\'{e}}rez-Hern{\'{a}}ndez}}, \bibinfo {author} {\bibfnamefont {A.~S.}\
  \bibnamefont {Landsman}}, \bibinfo {author} {\bibfnamefont {W.~A.}\
  \bibnamefont {Okell}}, \bibinfo {author} {\bibfnamefont {S.}~\bibnamefont
  {Zherebtsov}}, \bibinfo {author} {\bibfnamefont {B.}~\bibnamefont {Förg}},
  \bibinfo {author} {\bibfnamefont {J.}~\bibnamefont {Schötz}}, \bibinfo
  {author} {\bibfnamefont {L.}~\bibnamefont {Seiffert}}, \bibinfo {author}
  {\bibfnamefont {T.}~\bibnamefont {Fennel}}, \bibinfo {author} {\bibfnamefont
  {T.}~\bibnamefont {Shaaran}}, \bibinfo {author} {\bibfnamefont
  {T.}~\bibnamefont {Zimmermann}}, \bibinfo {author} {\bibfnamefont
  {A.}~\bibnamefont {Chac{\'{o}}n}}, \bibinfo {author} {\bibfnamefont
  {R.}~\bibnamefont {Guichard}}, \bibinfo {author} {\bibfnamefont
  {A.}~\bibnamefont {Zaïr}}, \bibinfo {author} {\bibfnamefont {J.~W.~G.}\
  \bibnamefont {Tisch}}, \bibinfo {author} {\bibfnamefont {J.~P.}\ \bibnamefont
  {Marangos}}, \bibinfo {author} {\bibfnamefont {T.}~\bibnamefont {Witting}},
  \bibinfo {author} {\bibfnamefont {A.}~\bibnamefont {Braun}}, \bibinfo
  {author} {\bibfnamefont {S.~A.}\ \bibnamefont {Maier}}, \bibinfo {author}
  {\bibfnamefont {L.}~\bibnamefont {Roso}}, \bibinfo {author} {\bibfnamefont
  {M.}~\bibnamefont {Krüger}}, \bibinfo {author} {\bibfnamefont
  {P.}~\bibnamefont {Hommelhoff}}, \bibinfo {author} {\bibfnamefont {M.~F.}\
  \bibnamefont {Kling}}, \bibinfo {author} {\bibfnamefont {F.}~\bibnamefont
  {Krausz}},\ and\ \bibinfo {author} {\bibfnamefont {M.}~\bibnamefont
  {Lewenstein}},\ }\bibfield  {title} {\bibinfo {title} {Attosecond physics at
  the nanoscale},\ }\href {https://doi.org/10.1088/1361-6633/aa574e} {\bibfield
   {journal} {\bibinfo  {journal} {Rep. Prog. Phys.}\ }\textbf {\bibinfo
  {volume} {80}},\ \bibinfo {pages} {054401} (\bibinfo {year}
  {2017})}\BibitemShut {NoStop}%
\bibitem [{\citenamefont {Dombi}\ \emph {et~al.}(2020)\citenamefont {Dombi},
  \citenamefont {P\'apa}, \citenamefont {Vogelsang}, \citenamefont {Yalunin},
  \citenamefont {Sivis}, \citenamefont {Herink}, \citenamefont {Sch\"afer},
  \citenamefont {Gro\ss{}}, \citenamefont {Ropers},\ and\ \citenamefont
  {Lienau}}]{DombiNano_2020}%
  \BibitemOpen
  \bibfield  {author} {\bibinfo {author} {\bibfnamefont {P.}~\bibnamefont
  {Dombi}}, \bibinfo {author} {\bibfnamefont {Z.}~\bibnamefont {P\'apa}},
  \bibinfo {author} {\bibfnamefont {J.}~\bibnamefont {Vogelsang}}, \bibinfo
  {author} {\bibfnamefont {S.~V.}\ \bibnamefont {Yalunin}}, \bibinfo {author}
  {\bibfnamefont {M.}~\bibnamefont {Sivis}}, \bibinfo {author} {\bibfnamefont
  {G.}~\bibnamefont {Herink}}, \bibinfo {author} {\bibfnamefont
  {S.}~\bibnamefont {Sch\"afer}}, \bibinfo {author} {\bibfnamefont
  {P.}~\bibnamefont {Gro\ss{}}}, \bibinfo {author} {\bibfnamefont
  {C.}~\bibnamefont {Ropers}},\ and\ \bibinfo {author} {\bibfnamefont
  {C.}~\bibnamefont {Lienau}},\ }\bibfield  {title} {\bibinfo {title}
  {Strong-field nano-optics},\ }\href
  {https://doi.org/10.1103/RevModPhys.92.025003} {\bibfield  {journal}
  {\bibinfo  {journal} {Rev. Mod. Phys.}\ }\textbf {\bibinfo {volume} {92}},\
  \bibinfo {pages} {025003} (\bibinfo {year} {2020})}\BibitemShut {NoStop}%
\bibitem [{\citenamefont {Haessler}\ \emph {et~al.}(2010)\citenamefont
  {Haessler}, \citenamefont {Caillat}, \citenamefont {Boutu}, \citenamefont
  {Giovanetti-Teixeira}, \citenamefont {Ruchon}, \citenamefont {Auguste},
  \citenamefont {Diveki}, \citenamefont {Breger}, \citenamefont {Maquet},
  \citenamefont {Carr{\'{e}}}, \citenamefont {Ta{\"{i}}eb},\ and\ \citenamefont
  {Sali{\`{e}}res}}]{Haessler2010}%
  \BibitemOpen
  \bibfield  {author} {\bibinfo {author} {\bibfnamefont {S.}~\bibnamefont
  {Haessler}}, \bibinfo {author} {\bibfnamefont {J.}~\bibnamefont {Caillat}},
  \bibinfo {author} {\bibfnamefont {W.}~\bibnamefont {Boutu}}, \bibinfo
  {author} {\bibfnamefont {C.}~\bibnamefont {Giovanetti-Teixeira}}, \bibinfo
  {author} {\bibfnamefont {T.}~\bibnamefont {Ruchon}}, \bibinfo {author}
  {\bibfnamefont {T.}~\bibnamefont {Auguste}}, \bibinfo {author} {\bibfnamefont
  {Z.}~\bibnamefont {Diveki}}, \bibinfo {author} {\bibfnamefont
  {P.}~\bibnamefont {Breger}}, \bibinfo {author} {\bibfnamefont
  {A.}~\bibnamefont {Maquet}}, \bibinfo {author} {\bibfnamefont
  {B.}~\bibnamefont {Carr{\'{e}}}}, \bibinfo {author} {\bibfnamefont
  {R.}~\bibnamefont {Ta{\"{i}}eb}},\ and\ \bibinfo {author} {\bibfnamefont
  {P.}~\bibnamefont {Sali{\`{e}}res}},\ }\bibfield  {title} {\bibinfo {title}
  {{Attosecond imaging of molecular electronic wavepackets}},\ }\href
  {https://doi.org/10.1038/nphys1511} {\bibfield  {journal} {\bibinfo
  {journal} {Nature Physics}\ }\textbf {\bibinfo {volume} {6}},\ \bibinfo
  {pages} {200} (\bibinfo {year} {2010})}\BibitemShut {NoStop}%
\bibitem [{\citenamefont {Goulielmakis}\ \emph {et~al.}(2008)\citenamefont
  {Goulielmakis}, \citenamefont {Schultze}, \citenamefont {Hofstetter},
  \citenamefont {Yakovlev}, \citenamefont {Gagnon}, \citenamefont {Uiberacker},
  \citenamefont {Aquila}, \citenamefont {Gullikson}, \citenamefont {Attwood},
  \citenamefont {Kienberger}, \citenamefont {Krausz},\ and\ \citenamefont
  {Kleineberg}}]{Goulielmakis2008}%
  \BibitemOpen
  \bibfield  {author} {\bibinfo {author} {\bibfnamefont {E.}~\bibnamefont
  {Goulielmakis}}, \bibinfo {author} {\bibfnamefont {M.}~\bibnamefont
  {Schultze}}, \bibinfo {author} {\bibfnamefont {M.}~\bibnamefont
  {Hofstetter}}, \bibinfo {author} {\bibfnamefont {V.~S.}\ \bibnamefont
  {Yakovlev}}, \bibinfo {author} {\bibfnamefont {J.}~\bibnamefont {Gagnon}},
  \bibinfo {author} {\bibfnamefont {M.}~\bibnamefont {Uiberacker}}, \bibinfo
  {author} {\bibfnamefont {A.~L.}\ \bibnamefont {Aquila}}, \bibinfo {author}
  {\bibfnamefont {E.~M.}\ \bibnamefont {Gullikson}}, \bibinfo {author}
  {\bibfnamefont {D.~T.}\ \bibnamefont {Attwood}}, \bibinfo {author}
  {\bibfnamefont {R.}~\bibnamefont {Kienberger}}, \bibinfo {author}
  {\bibfnamefont {F.}~\bibnamefont {Krausz}},\ and\ \bibinfo {author}
  {\bibfnamefont {U.}~\bibnamefont {Kleineberg}},\ }\bibfield  {title}
  {\bibinfo {title} {{Single-cycle nonlinear optics.}},\ }\href
  {https://doi.org/10.1126/science.1157846} {\bibfield  {journal} {\bibinfo
  {journal} {Science}\ }\textbf {\bibinfo {volume} {320}},\ \bibinfo {pages}
  {1614} (\bibinfo {year} {2008})}\BibitemShut {NoStop}%
\bibitem [{\citenamefont {Itatani}\ \emph {et~al.}(2002)\citenamefont
  {Itatani}, \citenamefont {Qu{\'{e}}r{\'{e}}}, \citenamefont {Yudin},
  \citenamefont {Ivanov}, \citenamefont {Krausz},\ and\ \citenamefont
  {Corkum}}]{Itatani2002}%
  \BibitemOpen
  \bibfield  {author} {\bibinfo {author} {\bibfnamefont {J.}~\bibnamefont
  {Itatani}}, \bibinfo {author} {\bibfnamefont {F.}~\bibnamefont
  {Qu{\'{e}}r{\'{e}}}}, \bibinfo {author} {\bibfnamefont {G.~L.}\ \bibnamefont
  {Yudin}}, \bibinfo {author} {\bibfnamefont {M.~Y.}\ \bibnamefont {Ivanov}},
  \bibinfo {author} {\bibfnamefont {F.}~\bibnamefont {Krausz}},\ and\ \bibinfo
  {author} {\bibfnamefont {P.~B.}\ \bibnamefont {Corkum}},\ }\bibfield  {title}
  {\bibinfo {title} {{Attosecond Streak Camera}},\ }\href
  {https://doi.org/10.1103/PhysRevLett.88.173903} {\bibfield  {journal}
  {\bibinfo  {journal} {Phys. Rev. Lett.}\ }\textbf {\bibinfo {volume} {88}},\
  \bibinfo {pages} {173903} (\bibinfo {year} {2002})}\BibitemShut {NoStop}%
\bibitem [{\citenamefont {Sansone}\ \emph {et~al.}(2006)\citenamefont
  {Sansone}, \citenamefont {Benedetti}, \citenamefont {Calegari}, \citenamefont
  {Vozzi}, \citenamefont {Avaldi}, \citenamefont {Flammini}, \citenamefont
  {Poletto}, \citenamefont {Villoresi}, \citenamefont {Altucci}, \citenamefont
  {Velotta}, \citenamefont {Stagira}, \citenamefont {{De Silvestri}},\ and\
  \citenamefont {Nisoli}}]{Villoresi2006}%
  \BibitemOpen
  \bibfield  {author} {\bibinfo {author} {\bibfnamefont {G.}~\bibnamefont
  {Sansone}}, \bibinfo {author} {\bibfnamefont {E.}~\bibnamefont {Benedetti}},
  \bibinfo {author} {\bibfnamefont {F.}~\bibnamefont {Calegari}}, \bibinfo
  {author} {\bibfnamefont {C.}~\bibnamefont {Vozzi}}, \bibinfo {author}
  {\bibfnamefont {L.}~\bibnamefont {Avaldi}}, \bibinfo {author} {\bibfnamefont
  {R.}~\bibnamefont {Flammini}}, \bibinfo {author} {\bibfnamefont
  {L.}~\bibnamefont {Poletto}}, \bibinfo {author} {\bibfnamefont
  {P.}~\bibnamefont {Villoresi}}, \bibinfo {author} {\bibfnamefont
  {C.}~\bibnamefont {Altucci}}, \bibinfo {author} {\bibfnamefont
  {R.}~\bibnamefont {Velotta}}, \bibinfo {author} {\bibfnamefont
  {S.}~\bibnamefont {Stagira}}, \bibinfo {author} {\bibfnamefont
  {S.}~\bibnamefont {{De Silvestri}}},\ and\ \bibinfo {author} {\bibfnamefont
  {M.}~\bibnamefont {Nisoli}},\ }\bibfield  {title} {\bibinfo {title}
  {{Isolated Single-Cycle Attosecond Pulses}},\ }\href
  {https://doi.org/10.1126/science.1132838} {\bibfield  {journal} {\bibinfo
  {journal} {Science}\ }\textbf {\bibinfo {volume} {314}},\ \bibinfo {pages}
  {443} (\bibinfo {year} {2006})}\BibitemShut {NoStop}%
\bibitem [{\citenamefont {Cavalieri}\ \emph {et~al.}(2007)\citenamefont
  {Cavalieri}, \citenamefont {M{\"{u}}ller}, \citenamefont {Uphues},
  \citenamefont {Yakovlev}, \citenamefont {Baltu{\v{s}}ka}, \citenamefont
  {Horvath}, \citenamefont {Schmidt}, \citenamefont {Bl{\"{u}}mel},
  \citenamefont {Holzwarth}, \citenamefont {Hendel}, \citenamefont {Drescher},
  \citenamefont {Kleineberg}, \citenamefont {Echenique}, \citenamefont
  {Kienberger}, \citenamefont {Krausz},\ and\ \citenamefont
  {Heinzmann}}]{Horvath2007}%
  \BibitemOpen
  \bibfield  {author} {\bibinfo {author} {\bibfnamefont {A.~L.}\ \bibnamefont
  {Cavalieri}}, \bibinfo {author} {\bibfnamefont {N.}~\bibnamefont
  {M{\"{u}}ller}}, \bibinfo {author} {\bibfnamefont {T.}~\bibnamefont
  {Uphues}}, \bibinfo {author} {\bibfnamefont {V.~S.}\ \bibnamefont
  {Yakovlev}}, \bibinfo {author} {\bibfnamefont {A.}~\bibnamefont
  {Baltu{\v{s}}ka}}, \bibinfo {author} {\bibfnamefont {B.}~\bibnamefont
  {Horvath}}, \bibinfo {author} {\bibfnamefont {B.}~\bibnamefont {Schmidt}},
  \bibinfo {author} {\bibfnamefont {L.}~\bibnamefont {Bl{\"{u}}mel}}, \bibinfo
  {author} {\bibfnamefont {R.}~\bibnamefont {Holzwarth}}, \bibinfo {author}
  {\bibfnamefont {S.}~\bibnamefont {Hendel}}, \bibinfo {author} {\bibfnamefont
  {M.}~\bibnamefont {Drescher}}, \bibinfo {author} {\bibfnamefont
  {U.}~\bibnamefont {Kleineberg}}, \bibinfo {author} {\bibfnamefont {P.~M.}\
  \bibnamefont {Echenique}}, \bibinfo {author} {\bibfnamefont {R.}~\bibnamefont
  {Kienberger}}, \bibinfo {author} {\bibfnamefont {F.}~\bibnamefont {Krausz}},\
  and\ \bibinfo {author} {\bibfnamefont {U.}~\bibnamefont {Heinzmann}},\
  }\bibfield  {title} {\bibinfo {title} {{Attosecond spectroscopy in condensed
  matter}},\ }\href {https://doi.org/10.1038/nature06229} {\bibfield  {journal}
  {\bibinfo  {journal} {Nature}\ }\textbf {\bibinfo {volume} {449}},\ \bibinfo
  {pages} {1029} (\bibinfo {year} {2007})}\BibitemShut {NoStop}%
\bibitem [{\citenamefont {Pabst}\ \emph {et~al.}(2016)\citenamefont {Pabst},
  \citenamefont {Lein},\ and\ \citenamefont {W\"orner}}]{SFI_Pabst}%
  \BibitemOpen
  \bibfield  {author} {\bibinfo {author} {\bibfnamefont {S.}~\bibnamefont
  {Pabst}}, \bibinfo {author} {\bibfnamefont {M.}~\bibnamefont {Lein}},\ and\
  \bibinfo {author} {\bibfnamefont {H.~J.}\ \bibnamefont {W\"orner}},\
  }\bibfield  {title} {\bibinfo {title} {Preparing attosecond coherences by
  strong-field ionization},\ }\href
  {https://doi.org/10.1103/PhysRevA.93.023412} {\bibfield  {journal} {\bibinfo
  {journal} {Phys. Rev. A}\ }\textbf {\bibinfo {volume} {93}},\ \bibinfo
  {pages} {023412} (\bibinfo {year} {2016})}\BibitemShut {NoStop}%
\bibitem [{\citenamefont {Bourassin-Bouchet}\ \emph {et~al.}(2020)\citenamefont
  {Bourassin-Bouchet}, \citenamefont {Barreau}, \citenamefont {Gruson},
  \citenamefont {Hergott}, \citenamefont {Qu\'er\'e}, \citenamefont
  {Sali\`eres},\ and\ \citenamefont {Ruchon}}]{DecoherencePRX}%
  \BibitemOpen
  \bibfield  {author} {\bibinfo {author} {\bibfnamefont {C.}~\bibnamefont
  {Bourassin-Bouchet}}, \bibinfo {author} {\bibfnamefont {L.}~\bibnamefont
  {Barreau}}, \bibinfo {author} {\bibfnamefont {V.}~\bibnamefont {Gruson}},
  \bibinfo {author} {\bibfnamefont {J.-F.}\ \bibnamefont {Hergott}}, \bibinfo
  {author} {\bibfnamefont {F.}~\bibnamefont {Qu\'er\'e}}, \bibinfo {author}
  {\bibfnamefont {P.}~\bibnamefont {Sali\`eres}},\ and\ \bibinfo {author}
  {\bibfnamefont {T.}~\bibnamefont {Ruchon}},\ }\bibfield  {title} {\bibinfo
  {title} {Quantifying decoherence in attosecond metrology},\ }\href
  {https://doi.org/10.1103/PhysRevX.10.031048} {\bibfield  {journal} {\bibinfo
  {journal} {Phys. Rev. X}\ }\textbf {\bibinfo {volume} {10}},\ \bibinfo
  {pages} {031048} (\bibinfo {year} {2020})}\BibitemShut {NoStop}%
\bibitem [{\citenamefont {Mehmood}\ \emph
  {et~al.}(2021{\natexlab{a}})\citenamefont {Mehmood}, \citenamefont
  {Lindroth},\ and\ \citenamefont {Argenti}}]{Mehmood2021}%
  \BibitemOpen
  \bibfield  {author} {\bibinfo {author} {\bibfnamefont {S.}~\bibnamefont
  {Mehmood}}, \bibinfo {author} {\bibfnamefont {E.}~\bibnamefont {Lindroth}},\
  and\ \bibinfo {author} {\bibfnamefont {L.}~\bibnamefont {Argenti}},\
  }\bibfield  {title} {\bibinfo {title} {Coherence control in helium-ion
  ensembles},\ }\href {https://doi.org/10.1103/PhysRevResearch.3.023233}
  {\bibfield  {journal} {\bibinfo  {journal} {Phys. Rev. Research}\ }\textbf
  {\bibinfo {volume} {3}},\ \bibinfo {pages} {023233} (\bibinfo {year}
  {2021}{\natexlab{a}})}\BibitemShut {NoStop}%
\bibitem [{\citenamefont {Ossiander}\ \emph {et~al.}(2017)\citenamefont
  {Ossiander}, \citenamefont {Siegrist}, \citenamefont {Shirvanyan},
  \citenamefont {Pazourek}, \citenamefont {Sommer}, \citenamefont {Latka},
  \citenamefont {Guggenmos}, \citenamefont {Nagele}, \citenamefont {Feist},
  \citenamefont {Burgd{\"{o}}rfer}, \citenamefont {Kienberger},\ and\
  \citenamefont {Schultze}}]{Ossiander2016}%
  \BibitemOpen
  \bibfield  {author} {\bibinfo {author} {\bibfnamefont {M.}~\bibnamefont
  {Ossiander}}, \bibinfo {author} {\bibfnamefont {F.}~\bibnamefont {Siegrist}},
  \bibinfo {author} {\bibfnamefont {V.}~\bibnamefont {Shirvanyan}}, \bibinfo
  {author} {\bibfnamefont {R.}~\bibnamefont {Pazourek}}, \bibinfo {author}
  {\bibfnamefont {A.}~\bibnamefont {Sommer}}, \bibinfo {author} {\bibfnamefont
  {T.}~\bibnamefont {Latka}}, \bibinfo {author} {\bibfnamefont
  {A.}~\bibnamefont {Guggenmos}}, \bibinfo {author} {\bibfnamefont
  {S.}~\bibnamefont {Nagele}}, \bibinfo {author} {\bibfnamefont
  {J.}~\bibnamefont {Feist}}, \bibinfo {author} {\bibfnamefont
  {J.}~\bibnamefont {Burgd{\"{o}}rfer}}, \bibinfo {author} {\bibfnamefont
  {R.}~\bibnamefont {Kienberger}},\ and\ \bibinfo {author} {\bibfnamefont
  {M.}~\bibnamefont {Schultze}},\ }\bibfield  {title} {\bibinfo {title}
  {Attosecond correlation dynamics},\ }\href
  {https://doi.org/10.1038/nphys3941} {\bibfield  {journal} {\bibinfo
  {journal} {Nature Physics}\ }\textbf {\bibinfo {volume} {13}},\ \bibinfo
  {pages} {280} (\bibinfo {year} {2017})}\BibitemShut {NoStop}%
\bibitem [{\citenamefont {Goulielmakis}\ \emph {et~al.}(2010)\citenamefont
  {Goulielmakis}, \citenamefont {Loh}, \citenamefont {Wirth}, \citenamefont
  {Santra}, \citenamefont {Rohringer}, \citenamefont {Yakovlev}, \citenamefont
  {Zherebtsov}, \citenamefont {Pfeifer}, \citenamefont {Azzeer}, \citenamefont
  {Kling}, \citenamefont {Leone},\ and\ \citenamefont
  {Krausz}}]{GoulielmakisNAT2010}%
  \BibitemOpen
  \bibfield  {author} {\bibinfo {author} {\bibfnamefont {E.}~\bibnamefont
  {Goulielmakis}}, \bibinfo {author} {\bibfnamefont {Z.-H.}\ \bibnamefont
  {Loh}}, \bibinfo {author} {\bibfnamefont {A.}~\bibnamefont {Wirth}}, \bibinfo
  {author} {\bibfnamefont {R.}~\bibnamefont {Santra}}, \bibinfo {author}
  {\bibfnamefont {N.}~\bibnamefont {Rohringer}}, \bibinfo {author}
  {\bibfnamefont {V.~S.}\ \bibnamefont {Yakovlev}}, \bibinfo {author}
  {\bibfnamefont {S.}~\bibnamefont {Zherebtsov}}, \bibinfo {author}
  {\bibfnamefont {T.}~\bibnamefont {Pfeifer}}, \bibinfo {author} {\bibfnamefont
  {A.~M.}\ \bibnamefont {Azzeer}}, \bibinfo {author} {\bibfnamefont {M.~F.}\
  \bibnamefont {Kling}}, \bibinfo {author} {\bibfnamefont {S.~R.}\ \bibnamefont
  {Leone}},\ and\ \bibinfo {author} {\bibfnamefont {F.}~\bibnamefont
  {Krausz}},\ }\bibfield  {title} {\bibinfo {title} {Real-time observation of
  valence electron motion},\ }\href {https://doi.org/10.1038/nature09212}
  {\bibfield  {journal} {\bibinfo  {journal} {Nature}\ }\textbf {\bibinfo
  {volume} {466}},\ \bibinfo {pages} {739} (\bibinfo {year}
  {2010})}\BibitemShut {NoStop}%
\bibitem [{\citenamefont {Guillemin}\ \emph {et~al.}(2015)\citenamefont
  {Guillemin}, \citenamefont {Decleva}, \citenamefont {Stener}, \citenamefont
  {Bomme}, \citenamefont {Marin}, \citenamefont {Journel}, \citenamefont
  {Marchenko}, \citenamefont {Kushawaha}, \citenamefont {J{\"a}nk{\"a}l{\"a}},
  \citenamefont {Trcera}, \citenamefont {Bowen}, \citenamefont {Lindle},
  \citenamefont {Piancastelli},\ and\ \citenamefont {Simon}}]{Guillemin2015}%
  \BibitemOpen
  \bibfield  {author} {\bibinfo {author} {\bibfnamefont {R.}~\bibnamefont
  {Guillemin}}, \bibinfo {author} {\bibfnamefont {P.}~\bibnamefont {Decleva}},
  \bibinfo {author} {\bibfnamefont {M.}~\bibnamefont {Stener}}, \bibinfo
  {author} {\bibfnamefont {C.}~\bibnamefont {Bomme}}, \bibinfo {author}
  {\bibfnamefont {T.}~\bibnamefont {Marin}}, \bibinfo {author} {\bibfnamefont
  {L.}~\bibnamefont {Journel}}, \bibinfo {author} {\bibfnamefont
  {T.}~\bibnamefont {Marchenko}}, \bibinfo {author} {\bibfnamefont {R.~K.}\
  \bibnamefont {Kushawaha}}, \bibinfo {author} {\bibfnamefont {K.}~\bibnamefont
  {J{\"a}nk{\"a}l{\"a}}}, \bibinfo {author} {\bibfnamefont {N.}~\bibnamefont
  {Trcera}}, \bibinfo {author} {\bibfnamefont {K.~P.}\ \bibnamefont {Bowen}},
  \bibinfo {author} {\bibfnamefont {D.~W.}\ \bibnamefont {Lindle}}, \bibinfo
  {author} {\bibfnamefont {M.~N.}\ \bibnamefont {Piancastelli}},\ and\ \bibinfo
  {author} {\bibfnamefont {M.}~\bibnamefont {Simon}},\ }\bibfield  {title}
  {\bibinfo {title} {Selecting core-hole localization or delocalization in
  cs$_2$ by photofragmentation dynamics},\ }\href
  {https://doi.org/10.1038/ncomms7166} {\bibfield  {journal} {\bibinfo
  {journal} {Nature Comm.}\ }\textbf {\bibinfo {volume} {6}},\ \bibinfo {pages}
  {6166} (\bibinfo {year} {2015})}\BibitemShut {NoStop}%
\bibitem [{\citenamefont {Shapiro}(2011)}]{ShapiroPRA2011}%
  \BibitemOpen
  \bibfield  {author} {\bibinfo {author} {\bibfnamefont {M.}~\bibnamefont
  {Shapiro}},\ }\bibfield  {title} {\bibinfo {title} {Generating and
  controlling chains of entangled atoms by coherent control techniques},\
  }\href {https://doi.org/10.1103/PhysRevA.84.053432} {\bibfield  {journal}
  {\bibinfo  {journal} {Phys. Rev. A}\ }\textbf {\bibinfo {volume} {84}},\
  \bibinfo {pages} {053432} (\bibinfo {year} {2011})}\BibitemShut {NoStop}%
\bibitem [{\citenamefont {Argenti}\ and\ \citenamefont
  {Lindroth}(2010)}]{Argenti2010}%
  \BibitemOpen
  \bibfield  {author} {\bibinfo {author} {\bibfnamefont {L.}~\bibnamefont
  {Argenti}}\ and\ \bibinfo {author} {\bibfnamefont {E.}~\bibnamefont
  {Lindroth}},\ }\bibfield  {title} {\bibinfo {title} {Ionization branching
  ratio control with a resonance attosecond clock},\ }\href
  {https://doi.org/10.1103/PhysRevLett.105.053002} {\bibfield  {journal}
  {\bibinfo  {journal} {Phys. Rev. Lett.}\ }\textbf {\bibinfo {volume} {105}},\
  \bibinfo {pages} {053002} (\bibinfo {year} {2010})}\BibitemShut {NoStop}%
\bibitem [{\citenamefont {Gruson}\ \emph {et~al.}(2016)\citenamefont {Gruson},
  \citenamefont {Barreau}, \citenamefont {Jim{\'e}nez-Galan}, \citenamefont
  {Risoud}, \citenamefont {Caillat}, \citenamefont {Maquet}, \citenamefont
  {Carr{\'e}}, \citenamefont {Lepetit}, \citenamefont {Hergott}, \citenamefont
  {Ruchon}, \citenamefont {Argenti}, \citenamefont {Ta{\"\i}eb}, \citenamefont
  {Mart{\'\i}n},\ and\ \citenamefont {Sali{\`e}res}}]{ArgentiScience}%
  \BibitemOpen
  \bibfield  {author} {\bibinfo {author} {\bibfnamefont {V.}~\bibnamefont
  {Gruson}}, \bibinfo {author} {\bibfnamefont {L.}~\bibnamefont {Barreau}},
  \bibinfo {author} {\bibfnamefont {{\'A}.}~\bibnamefont {Jim{\'e}nez-Galan}},
  \bibinfo {author} {\bibfnamefont {F.}~\bibnamefont {Risoud}}, \bibinfo
  {author} {\bibfnamefont {J.}~\bibnamefont {Caillat}}, \bibinfo {author}
  {\bibfnamefont {A.}~\bibnamefont {Maquet}}, \bibinfo {author} {\bibfnamefont
  {B.}~\bibnamefont {Carr{\'e}}}, \bibinfo {author} {\bibfnamefont
  {F.}~\bibnamefont {Lepetit}}, \bibinfo {author} {\bibfnamefont {J.-F.}\
  \bibnamefont {Hergott}}, \bibinfo {author} {\bibfnamefont {T.}~\bibnamefont
  {Ruchon}}, \bibinfo {author} {\bibfnamefont {L.}~\bibnamefont {Argenti}},
  \bibinfo {author} {\bibfnamefont {R.}~\bibnamefont {Ta{\"\i}eb}}, \bibinfo
  {author} {\bibfnamefont {F.}~\bibnamefont {Mart{\'\i}n}},\ and\ \bibinfo
  {author} {\bibfnamefont {P.}~\bibnamefont {Sali{\`e}res}},\ }\bibfield
  {title} {\bibinfo {title} {Attosecond dynamics through a fano resonance:
  Monitoring the birth of a photoelectron},\ }\href
  {https://doi.org/10.1126/science.aah5188} {\bibfield  {journal} {\bibinfo
  {journal} {Science}\ }\textbf {\bibinfo {volume} {354}},\ \bibinfo {pages}
  {734} (\bibinfo {year} {2016})}\BibitemShut {NoStop}%
\bibitem [{\citenamefont {Lindroth}\ and\ \citenamefont
  {Argenti}(2012)}]{EvaLuca_ChemRev}%
  \BibitemOpen
  \bibfield  {author} {\bibinfo {author} {\bibfnamefont {E.}~\bibnamefont
  {Lindroth}}\ and\ \bibinfo {author} {\bibfnamefont {L.}~\bibnamefont
  {Argenti}},\ }\bibfield  {title} {\bibinfo {title} {Chapter 5 - atomic
  resonance states and their role in charge-changing processes},\ }in\ \href
  {https://doi.org/https://doi.org/10.1016/B978-0-12-397009-1.00005-9} {\emph
  {\bibinfo {booktitle} {Adv. Quant. Chem.}}},\ \bibinfo {series} {Adv. Quant.
  Chem.}, Vol.~\bibinfo {volume} {63},\ \bibinfo {editor} {edited by\ \bibinfo
  {editor} {\bibfnamefont {C.~A.}\ \bibnamefont {Nicolaides}}, \bibinfo
  {editor} {\bibfnamefont {E.}~\bibnamefont {Brändas}},\ and\ \bibinfo
  {editor} {\bibfnamefont {J.~R.}\ \bibnamefont {Sabin}}}\ (\bibinfo
  {publisher} {Academic Press},\ \bibinfo {year} {2012})\ pp.\ \bibinfo {pages}
  {247 -- 308}\BibitemShut {NoStop}%
\bibitem [{\citenamefont {Jim\'enez-Gal\'an}\ \emph {et~al.}(2014)\citenamefont
  {Jim\'enez-Gal\'an}, \citenamefont {Argenti},\ and\ \citenamefont
  {Mart\'{\i}n}}]{LucaPRL2014}%
  \BibitemOpen
  \bibfield  {author} {\bibinfo {author} {\bibfnamefont {A.}~\bibnamefont
  {Jim\'enez-Gal\'an}}, \bibinfo {author} {\bibfnamefont {L.}~\bibnamefont
  {Argenti}},\ and\ \bibinfo {author} {\bibfnamefont {F.}~\bibnamefont
  {Mart\'{\i}n}},\ }\bibfield  {title} {\bibinfo {title} {Modulation of
  attosecond beating in resonant two-photon ionization},\ }\href
  {https://doi.org/10.1103/PhysRevLett.113.263001} {\bibfield  {journal}
  {\bibinfo  {journal} {Phys. Rev. Lett.}\ }\textbf {\bibinfo {volume} {113}},\
  \bibinfo {pages} {263001} (\bibinfo {year} {2014})}\BibitemShut {NoStop}%
\bibitem [{\citenamefont {Kotur}\ \emph {et~al.}(2016)\citenamefont {Kotur},
  \citenamefont {Gu{\'{e}}not}, \citenamefont {Jim{\'{e}}nez-Gal{\'{a}}n},
  \citenamefont {Kroon}, \citenamefont {Larsen}, \citenamefont {Louisy},
  \citenamefont {Bengtsson}, \citenamefont {Miranda}, \citenamefont
  {Mauritsson}, \citenamefont {Arnold}, \citenamefont {Canton}, \citenamefont
  {Gisselbrecht}, \citenamefont {Carette}, \citenamefont {Dahlstr{\"{o}}m},
  \citenamefont {Lindroth}, \citenamefont {Maquet}, \citenamefont {Argenti},
  \citenamefont {Mart{\'{i}}n},\ and\ \citenamefont {L'Huillier}}]{Kotur2016}%
  \BibitemOpen
  \bibfield  {author} {\bibinfo {author} {\bibfnamefont {M.}~\bibnamefont
  {Kotur}}, \bibinfo {author} {\bibfnamefont {D.}~\bibnamefont {Gu{\'{e}}not}},
  \bibinfo {author} {\bibfnamefont {{\'{A}}.}~\bibnamefont
  {Jim{\'{e}}nez-Gal{\'{a}}n}}, \bibinfo {author} {\bibfnamefont
  {D.}~\bibnamefont {Kroon}}, \bibinfo {author} {\bibfnamefont {E.~W.}\
  \bibnamefont {Larsen}}, \bibinfo {author} {\bibfnamefont {M.}~\bibnamefont
  {Louisy}}, \bibinfo {author} {\bibfnamefont {S.}~\bibnamefont {Bengtsson}},
  \bibinfo {author} {\bibfnamefont {M.}~\bibnamefont {Miranda}}, \bibinfo
  {author} {\bibfnamefont {J.}~\bibnamefont {Mauritsson}}, \bibinfo {author}
  {\bibfnamefont {C.~L.}\ \bibnamefont {Arnold}}, \bibinfo {author}
  {\bibfnamefont {S.~E.}\ \bibnamefont {Canton}}, \bibinfo {author}
  {\bibfnamefont {M.}~\bibnamefont {Gisselbrecht}}, \bibinfo {author}
  {\bibfnamefont {T.}~\bibnamefont {Carette}}, \bibinfo {author} {\bibfnamefont
  {J.~M.}\ \bibnamefont {Dahlstr{\"{o}}m}}, \bibinfo {author} {\bibfnamefont
  {E.}~\bibnamefont {Lindroth}}, \bibinfo {author} {\bibfnamefont
  {A.}~\bibnamefont {Maquet}}, \bibinfo {author} {\bibfnamefont
  {L.}~\bibnamefont {Argenti}}, \bibinfo {author} {\bibfnamefont
  {F.}~\bibnamefont {Mart{\'{i}}n}},\ and\ \bibinfo {author} {\bibfnamefont
  {A.}~\bibnamefont {L'Huillier}},\ }\bibfield  {title} {\bibinfo {title}
  {{Spectral phase measurement of a Fano resonance using tunable attosecond
  pulses}},\ }\href {https://doi.org/10.1038/ncomms10566} {\bibfield  {journal}
  {\bibinfo  {journal} {Nat. Commun.}\ }\textbf {\bibinfo {volume} {7}},\
  \bibinfo {pages} {10566} (\bibinfo {year} {2016})}\BibitemShut {NoStop}%
\bibitem [{\citenamefont {Ott}\ \emph {et~al.}(2014)\citenamefont {Ott},
  \citenamefont {Kaldun}, \citenamefont {Argenti}, \citenamefont {Raith},
  \citenamefont {Meyer}, \citenamefont {Laux}, \citenamefont {Zhang},
  \citenamefont {Bl{\"{a}}ttermann}, \citenamefont {Hagstotz}, \citenamefont
  {Ding}, \citenamefont {Heck}, \citenamefont {Madro{\~{n}}ero}, \citenamefont
  {Mart{\'{i}}n},\ and\ \citenamefont {Pfeifer}}]{Ott2014}%
  \BibitemOpen
  \bibfield  {author} {\bibinfo {author} {\bibfnamefont {C.}~\bibnamefont
  {Ott}}, \bibinfo {author} {\bibfnamefont {A.}~\bibnamefont {Kaldun}},
  \bibinfo {author} {\bibfnamefont {L.}~\bibnamefont {Argenti}}, \bibinfo
  {author} {\bibfnamefont {P.}~\bibnamefont {Raith}}, \bibinfo {author}
  {\bibfnamefont {K.}~\bibnamefont {Meyer}}, \bibinfo {author} {\bibfnamefont
  {M.}~\bibnamefont {Laux}}, \bibinfo {author} {\bibfnamefont {Y.}~\bibnamefont
  {Zhang}}, \bibinfo {author} {\bibfnamefont {A.}~\bibnamefont
  {Bl{\"{a}}ttermann}}, \bibinfo {author} {\bibfnamefont {S.}~\bibnamefont
  {Hagstotz}}, \bibinfo {author} {\bibfnamefont {T.}~\bibnamefont {Ding}},
  \bibinfo {author} {\bibfnamefont {R.}~\bibnamefont {Heck}}, \bibinfo {author}
  {\bibfnamefont {J.}~\bibnamefont {Madro{\~{n}}ero}}, \bibinfo {author}
  {\bibfnamefont {F.}~\bibnamefont {Mart{\'{i}}n}},\ and\ \bibinfo {author}
  {\bibfnamefont {T.}~\bibnamefont {Pfeifer}},\ }\bibfield  {title} {\bibinfo
  {title} {{Reconstruction and control of a time-dependent two-electron wave
  packet}},\ }\href {https://doi.org/10.1038/nature14026} {\bibfield  {journal}
  {\bibinfo  {journal} {Nature}\ }\textbf {\bibinfo {volume} {516}},\ \bibinfo
  {pages} {374} (\bibinfo {year} {2014})}\BibitemShut {NoStop}%
\bibitem [{\citenamefont {Nagasono}\ \emph {et~al.}(2007)\citenamefont
  {Nagasono}, \citenamefont {Suljoti}, \citenamefont {Pietzsch}, \citenamefont
  {Hennies}, \citenamefont {Wellh\"ofer}, \citenamefont {Hoeft}, \citenamefont
  {Martins}, \citenamefont {Wurth}, \citenamefont {Treusch}, \citenamefont
  {Feldhaus}, \citenamefont {Schneider},\ and\ \citenamefont
  {F\"ohlisch}}]{Nagasono2007}%
  \BibitemOpen
  \bibfield  {author} {\bibinfo {author} {\bibfnamefont {M.}~\bibnamefont
  {Nagasono}}, \bibinfo {author} {\bibfnamefont {E.}~\bibnamefont {Suljoti}},
  \bibinfo {author} {\bibfnamefont {A.}~\bibnamefont {Pietzsch}}, \bibinfo
  {author} {\bibfnamefont {F.}~\bibnamefont {Hennies}}, \bibinfo {author}
  {\bibfnamefont {M.}~\bibnamefont {Wellh\"ofer}}, \bibinfo {author}
  {\bibfnamefont {J.-T.}\ \bibnamefont {Hoeft}}, \bibinfo {author}
  {\bibfnamefont {M.}~\bibnamefont {Martins}}, \bibinfo {author} {\bibfnamefont
  {W.}~\bibnamefont {Wurth}}, \bibinfo {author} {\bibfnamefont
  {R.}~\bibnamefont {Treusch}}, \bibinfo {author} {\bibfnamefont
  {J.}~\bibnamefont {Feldhaus}}, \bibinfo {author} {\bibfnamefont {J.~R.}\
  \bibnamefont {Schneider}},\ and\ \bibinfo {author} {\bibfnamefont
  {A.}~\bibnamefont {F\"ohlisch}},\ }\bibfield  {title} {\bibinfo {title}
  {Resonant two-photon absorption of extreme-ultraviolet free-electron-laser
  radiation in helium},\ }\href {https://doi.org/10.1103/PhysRevA.75.051406}
  {\bibfield  {journal} {\bibinfo  {journal} {Phys. Rev. A}\ }\textbf {\bibinfo
  {volume} {75}},\ \bibinfo {pages} {051406} (\bibinfo {year}
  {2007})}\BibitemShut {NoStop}%
\bibitem [{\citenamefont {H{\"u}tten}\ \emph {et~al.}(2018)\citenamefont
  {H{\"u}tten}, \citenamefont {Mittermair}, \citenamefont {Stock},
  \citenamefont {Beerwerth}, \citenamefont {Shirvanyan}, \citenamefont
  {Riemensberger}, \citenamefont {Duensing}, \citenamefont {Heider},
  \citenamefont {Wagner}, \citenamefont {Guggenmos}, \citenamefont {Fritzsche},
  \citenamefont {Kabachnik}, \citenamefont {Kienberger},\ and\ \citenamefont
  {Bernhardt}}]{Htten2018}%
  \BibitemOpen
  \bibfield  {author} {\bibinfo {author} {\bibfnamefont {K.}~\bibnamefont
  {H{\"u}tten}}, \bibinfo {author} {\bibfnamefont {M.}~\bibnamefont
  {Mittermair}}, \bibinfo {author} {\bibfnamefont {S.~O.}\ \bibnamefont
  {Stock}}, \bibinfo {author} {\bibfnamefont {R.}~\bibnamefont {Beerwerth}},
  \bibinfo {author} {\bibfnamefont {V.}~\bibnamefont {Shirvanyan}}, \bibinfo
  {author} {\bibfnamefont {J.}~\bibnamefont {Riemensberger}}, \bibinfo {author}
  {\bibfnamefont {A.}~\bibnamefont {Duensing}}, \bibinfo {author}
  {\bibfnamefont {R.}~\bibnamefont {Heider}}, \bibinfo {author} {\bibfnamefont
  {M.~S.}\ \bibnamefont {Wagner}}, \bibinfo {author} {\bibfnamefont
  {A.}~\bibnamefont {Guggenmos}}, \bibinfo {author} {\bibfnamefont
  {S.}~\bibnamefont {Fritzsche}}, \bibinfo {author} {\bibfnamefont {N.~M.}\
  \bibnamefont {Kabachnik}}, \bibinfo {author} {\bibfnamefont {R.}~\bibnamefont
  {Kienberger}},\ and\ \bibinfo {author} {\bibfnamefont {B.}~\bibnamefont
  {Bernhardt}},\ }\bibfield  {title} {\bibinfo {title} {Ultrafast quantum
  control of ionization dynamics in krypton},\ }\href
  {https://doi.org/10.1038/s41467-018-03122-1} {\bibfield  {journal} {\bibinfo
  {journal} {Nature Comm.}\ }\textbf {\bibinfo {volume} {9}},\ \bibinfo {pages}
  {719} (\bibinfo {year} {2018})}\BibitemShut {NoStop}%
\bibitem [{\citenamefont {F{\"{o}}hlisch}\ \emph {et~al.}(2005)\citenamefont
  {F{\"{o}}hlisch}, \citenamefont {Feulner}, \citenamefont {Hennies},
  \citenamefont {Fink}, \citenamefont {Menzel}, \citenamefont {Sanchez-Portal},
  \citenamefont {Echenique},\ and\ \citenamefont {Wurth}}]{Fohlisch2005}%
  \BibitemOpen
  \bibfield  {author} {\bibinfo {author} {\bibfnamefont {a.}~\bibnamefont
  {F{\"{o}}hlisch}}, \bibinfo {author} {\bibfnamefont {P.}~\bibnamefont
  {Feulner}}, \bibinfo {author} {\bibfnamefont {F.}~\bibnamefont {Hennies}},
  \bibinfo {author} {\bibfnamefont {A.}~\bibnamefont {Fink}}, \bibinfo {author}
  {\bibfnamefont {D.}~\bibnamefont {Menzel}}, \bibinfo {author} {\bibfnamefont
  {D.}~\bibnamefont {Sanchez-Portal}}, \bibinfo {author} {\bibfnamefont
  {P.~M.}\ \bibnamefont {Echenique}},\ and\ \bibinfo {author} {\bibfnamefont
  {W.}~\bibnamefont {Wurth}},\ }\bibfield  {title} {\bibinfo {title} {{Direct
  observation of electron dynamics in the attosecond domain}},\ }\href
  {https://doi.org/10.1038/nature03833} {\bibfield  {journal} {\bibinfo
  {journal} {Nature}\ }\textbf {\bibinfo {volume} {436}},\ \bibinfo {pages}
  {373} (\bibinfo {year} {2005})}\BibitemShut {NoStop}%
\bibitem [{\citenamefont {Sansone}\ \emph {et~al.}(2010)\citenamefont
  {Sansone}, \citenamefont {Kelkensberg}, \citenamefont {P{\'{e}}rez-Torres},
  \citenamefont {Morales}, \citenamefont {Kling}, \citenamefont {Siu},
  \citenamefont {Ghafur}, \citenamefont {Johnsson}, \citenamefont {Swoboda},
  \citenamefont {Benedetti}, \citenamefont {Ferrari}, \citenamefont
  {L{\'{e}}pine}, \citenamefont {Sanz-Vicario}, \citenamefont {Zherebtsov},
  \citenamefont {Znakovskaya}, \citenamefont {L'Huillier}, \citenamefont
  {Ivanov}, \citenamefont {Nisoli}, \citenamefont {Mart{\'{i}}n},\ and\
  \citenamefont {Vrakking}}]{Sansone2010}%
  \BibitemOpen
  \bibfield  {author} {\bibinfo {author} {\bibfnamefont {G.}~\bibnamefont
  {Sansone}}, \bibinfo {author} {\bibfnamefont {F.}~\bibnamefont
  {Kelkensberg}}, \bibinfo {author} {\bibfnamefont {J.~F.}\ \bibnamefont
  {P{\'{e}}rez-Torres}}, \bibinfo {author} {\bibfnamefont {F.}~\bibnamefont
  {Morales}}, \bibinfo {author} {\bibfnamefont {M.~F.}\ \bibnamefont {Kling}},
  \bibinfo {author} {\bibfnamefont {W.}~\bibnamefont {Siu}}, \bibinfo {author}
  {\bibfnamefont {O.}~\bibnamefont {Ghafur}}, \bibinfo {author} {\bibfnamefont
  {P.}~\bibnamefont {Johnsson}}, \bibinfo {author} {\bibfnamefont
  {M.}~\bibnamefont {Swoboda}}, \bibinfo {author} {\bibfnamefont
  {E.}~\bibnamefont {Benedetti}}, \bibinfo {author} {\bibfnamefont
  {F.}~\bibnamefont {Ferrari}}, \bibinfo {author} {\bibfnamefont
  {F.}~\bibnamefont {L{\'{e}}pine}}, \bibinfo {author} {\bibfnamefont {J.~L.}\
  \bibnamefont {Sanz-Vicario}}, \bibinfo {author} {\bibfnamefont
  {S.}~\bibnamefont {Zherebtsov}}, \bibinfo {author} {\bibfnamefont
  {I.}~\bibnamefont {Znakovskaya}}, \bibinfo {author} {\bibfnamefont
  {A.}~\bibnamefont {L'Huillier}}, \bibinfo {author} {\bibfnamefont {M.~Y.}\
  \bibnamefont {Ivanov}}, \bibinfo {author} {\bibfnamefont {M.}~\bibnamefont
  {Nisoli}}, \bibinfo {author} {\bibfnamefont {F.}~\bibnamefont
  {Mart{\'{i}}n}},\ and\ \bibinfo {author} {\bibfnamefont {M.~J.~J.}\
  \bibnamefont {Vrakking}},\ }\bibfield  {title} {\bibinfo {title} {{Electron
  localization following attosecond molecular photoionization}},\ }\href
  {https://doi.org/10.1038/nature09084} {\bibfield  {journal} {\bibinfo
  {journal} {Nature}\ }\textbf {\bibinfo {volume} {465}},\ \bibinfo {pages}
  {763} (\bibinfo {year} {2010})}\BibitemShut {NoStop}%
\bibitem [{\citenamefont {Martin}\ \emph {et~al.}(2007)\citenamefont {Martin},
  \citenamefont {Fernandez}, \citenamefont {Havermeier}, \citenamefont
  {Foucar}, \citenamefont {Weber}, \citenamefont {Kreidi}, \citenamefont
  {Schoffler}, \citenamefont {Schmidt}, \citenamefont {Jahnke}, \citenamefont
  {Jagutzki}, \citenamefont {Czasch}, \citenamefont {Benis}, \citenamefont
  {Osipov}, \citenamefont {Landers}, \citenamefont {Belkacem}, \citenamefont
  {Prior}, \citenamefont {Schmidt-Bocking}, \citenamefont {Cocke},\ and\
  \citenamefont {Dorner}}]{Martin2007}%
  \BibitemOpen
  \bibfield  {author} {\bibinfo {author} {\bibfnamefont {F.}~\bibnamefont
  {Martin}}, \bibinfo {author} {\bibfnamefont {J.}~\bibnamefont {Fernandez}},
  \bibinfo {author} {\bibfnamefont {T.}~\bibnamefont {Havermeier}}, \bibinfo
  {author} {\bibfnamefont {L.}~\bibnamefont {Foucar}}, \bibinfo {author}
  {\bibfnamefont {T.}~\bibnamefont {Weber}}, \bibinfo {author} {\bibfnamefont
  {K.}~\bibnamefont {Kreidi}}, \bibinfo {author} {\bibfnamefont
  {M.}~\bibnamefont {Schoffler}}, \bibinfo {author} {\bibfnamefont
  {L.}~\bibnamefont {Schmidt}}, \bibinfo {author} {\bibfnamefont
  {T.}~\bibnamefont {Jahnke}}, \bibinfo {author} {\bibfnamefont
  {O.}~\bibnamefont {Jagutzki}}, \bibinfo {author} {\bibfnamefont
  {A.}~\bibnamefont {Czasch}}, \bibinfo {author} {\bibfnamefont {E.~P.}\
  \bibnamefont {Benis}}, \bibinfo {author} {\bibfnamefont {T.}~\bibnamefont
  {Osipov}}, \bibinfo {author} {\bibfnamefont {a.~L.}\ \bibnamefont {Landers}},
  \bibinfo {author} {\bibfnamefont {A.}~\bibnamefont {Belkacem}}, \bibinfo
  {author} {\bibfnamefont {M.~H.}\ \bibnamefont {Prior}}, \bibinfo {author}
  {\bibfnamefont {H.}~\bibnamefont {Schmidt-Bocking}}, \bibinfo {author}
  {\bibfnamefont {C.~L.}\ \bibnamefont {Cocke}},\ and\ \bibinfo {author}
  {\bibfnamefont {R.}~\bibnamefont {Dorner}},\ }\bibfield  {title} {\bibinfo
  {title} {{Single Photon-Induced Symmetry Breaking of H2 Dissociation}},\
  }\href {https://doi.org/10.1126/science.1136598} {\bibfield  {journal}
  {\bibinfo  {journal} {Science}\ }\textbf {\bibinfo {volume} {315}},\ \bibinfo
  {pages} {629} (\bibinfo {year} {2007})}\BibitemShut {NoStop}%
\bibitem [{\citenamefont {Doughty}\ \emph {et~al.}(2011)\citenamefont
  {Doughty}, \citenamefont {Haber}, \citenamefont {Hackett},\ and\
  \citenamefont {Leone}}]{Doughty2011}%
  \BibitemOpen
  \bibfield  {author} {\bibinfo {author} {\bibfnamefont {B.}~\bibnamefont
  {Doughty}}, \bibinfo {author} {\bibfnamefont {L.~H.}\ \bibnamefont {Haber}},
  \bibinfo {author} {\bibfnamefont {C.}~\bibnamefont {Hackett}},\ and\ \bibinfo
  {author} {\bibfnamefont {S.~R.}\ \bibnamefont {Leone}},\ }\bibfield  {title}
  {\bibinfo {title} {{Photoelectron angular distributions from autoionizing
  4s\textsuperscript{1}4p\textsuperscript{6}6p\textsuperscript{1} states in
  atomic krypton probed with femtosecond time resolution}},\ }\href
  {https://doi.org/10.1063/1.3547459} {\bibfield  {journal} {\bibinfo
  {journal} {J. Chem. Phys.}\ }\textbf {\bibinfo {volume} {134}},\ \bibinfo
  {pages} {094307} (\bibinfo {year} {2011})}\BibitemShut {NoStop}%
\bibitem [{\citenamefont {Wickenhauser}\ \emph {et~al.}(2006)\citenamefont
  {Wickenhauser}, \citenamefont {Burgd{\"{o}}rfer}, \citenamefont {Krausz},\
  and\ \citenamefont {Drescher}}]{Wickenhauser2006}%
  \BibitemOpen
  \bibfield  {author} {\bibinfo {author} {\bibfnamefont {M.}~\bibnamefont
  {Wickenhauser}}, \bibinfo {author} {\bibfnamefont {J.}~\bibnamefont
  {Burgd{\"{o}}rfer}}, \bibinfo {author} {\bibfnamefont {F.}~\bibnamefont
  {Krausz}},\ and\ \bibinfo {author} {\bibfnamefont {M.}~\bibnamefont
  {Drescher}},\ }\bibfield  {title} {\bibinfo {title} {{Attosecond streaking of
  overlapping Fano resonances}},\ }\href
  {https://doi.org/10.1080/09500340500259870} {\bibfield  {journal} {\bibinfo
  {journal} {J. Mod. Opt.}\ }\textbf {\bibinfo {volume} {53-1}},\ \bibinfo
  {pages} {247} (\bibinfo {year} {2006})}\BibitemShut {NoStop}%
\bibitem [{\citenamefont {Cirelli}\ \emph {et~al.}(2018)\citenamefont
  {Cirelli}, \citenamefont {Marante}, \citenamefont {Heuser}, \citenamefont
  {Petersson}, \citenamefont {Jim{\'{e}}nez-Gal{\'{a}}n}, \citenamefont
  {Argenti}, \citenamefont {Zhong}, \citenamefont {Busto}, \citenamefont
  {Isinger}, \citenamefont {Nandi}, \citenamefont {Maclot}, \citenamefont
  {Rading}, \citenamefont {Johnsson}, \citenamefont {Gisselbrecht},
  \citenamefont {Lucchini}, \citenamefont {Gallmann}, \citenamefont
  {Dahlstr{\"{o}}m}, \citenamefont {Lindroth}, \citenamefont {L'Huillier},
  \citenamefont {Mart{\'{i}}n},\ and\ \citenamefont {Keller}}]{Cirelli2018}%
  \BibitemOpen
  \bibfield  {author} {\bibinfo {author} {\bibfnamefont {C.}~\bibnamefont
  {Cirelli}}, \bibinfo {author} {\bibfnamefont {C.}~\bibnamefont {Marante}},
  \bibinfo {author} {\bibfnamefont {S.}~\bibnamefont {Heuser}}, \bibinfo
  {author} {\bibfnamefont {C.~L.~M.}\ \bibnamefont {Petersson}}, \bibinfo
  {author} {\bibfnamefont {{\'{A}}.}~\bibnamefont {Jim{\'{e}}nez-Gal{\'{a}}n}},
  \bibinfo {author} {\bibfnamefont {L.}~\bibnamefont {Argenti}}, \bibinfo
  {author} {\bibfnamefont {S.}~\bibnamefont {Zhong}}, \bibinfo {author}
  {\bibfnamefont {D.}~\bibnamefont {Busto}}, \bibinfo {author} {\bibfnamefont
  {M.}~\bibnamefont {Isinger}}, \bibinfo {author} {\bibfnamefont
  {S.}~\bibnamefont {Nandi}}, \bibinfo {author} {\bibfnamefont
  {S.}~\bibnamefont {Maclot}}, \bibinfo {author} {\bibfnamefont
  {L.}~\bibnamefont {Rading}}, \bibinfo {author} {\bibfnamefont
  {P.}~\bibnamefont {Johnsson}}, \bibinfo {author} {\bibfnamefont
  {M.}~\bibnamefont {Gisselbrecht}}, \bibinfo {author} {\bibfnamefont
  {M.}~\bibnamefont {Lucchini}}, \bibinfo {author} {\bibfnamefont
  {L.}~\bibnamefont {Gallmann}}, \bibinfo {author} {\bibfnamefont {J.~M.}\
  \bibnamefont {Dahlstr{\"{o}}m}}, \bibinfo {author} {\bibfnamefont
  {E.}~\bibnamefont {Lindroth}}, \bibinfo {author} {\bibfnamefont
  {A.}~\bibnamefont {L'Huillier}}, \bibinfo {author} {\bibfnamefont
  {F.}~\bibnamefont {Mart{\'{i}}n}},\ and\ \bibinfo {author} {\bibfnamefont
  {U.}~\bibnamefont {Keller}},\ }\bibfield  {title} {\bibinfo {title}
  {{Anisotropic photoemission time delays close to a Fano resonance}},\ }\href
  {https://doi.org/10.1038/s41467-018-03009-1} {\bibfield  {journal} {\bibinfo
  {journal} {Nat. Commun.}\ }\textbf {\bibinfo {volume} {9}},\ \bibinfo {pages}
  {955} (\bibinfo {year} {2018})}\BibitemShut {NoStop}%
\bibitem [{\citenamefont {Drake}(2006)}]{bookSpringer}%
  \BibitemOpen
  \bibfield  {author} {\bibinfo {author} {\bibfnamefont {G.}~\bibnamefont
  {Drake}},\ }\href {https://doi.org/10.1007/978-0-387-26308-3} {\emph
  {\bibinfo {title} {Springer Handbook of Atomic, Molecular, and Optical
  Physics}}}\ (\bibinfo {year} {2006})\BibitemShut {NoStop}%
\bibitem [{\citenamefont {Argenti}\ and\ \citenamefont
  {Moccia}(2006)}]{Argenti2006}%
  \BibitemOpen
  \bibfield  {author} {\bibinfo {author} {\bibfnamefont {L.}~\bibnamefont
  {Argenti}}\ and\ \bibinfo {author} {\bibfnamefont {R.}~\bibnamefont
  {Moccia}},\ }\bibfield  {title} {\bibinfo {title} {{K-matrix method with
  B-splines : $\sigma_N$, $\beta_N$ and resonances in He photoionization below
  $N=4$ threshold}},\ }\href {https://doi.org/10.1088/0953-4075/39/12/012}
  {\bibfield  {journal} {\bibinfo  {journal} {J. Phys. B: At. Mol. Opt. Phys.}\
  }\textbf {\bibinfo {volume} {39}},\ \bibinfo {pages} {2773} (\bibinfo {year}
  {2006})}\BibitemShut {NoStop}%
\bibitem [{\citenamefont {Carette}\ \emph {et~al.}(2013)\citenamefont
  {Carette}, \citenamefont {Dahlstr{\"{o}}m}, \citenamefont {Argenti},\ and\
  \citenamefont {Lindroth}}]{Carette2013}%
  \BibitemOpen
  \bibfield  {author} {\bibinfo {author} {\bibfnamefont {T.}~\bibnamefont
  {Carette}}, \bibinfo {author} {\bibfnamefont {J.~M.}\ \bibnamefont
  {Dahlstr{\"{o}}m}}, \bibinfo {author} {\bibfnamefont {L.}~\bibnamefont
  {Argenti}},\ and\ \bibinfo {author} {\bibfnamefont {E.}~\bibnamefont
  {Lindroth}},\ }\bibfield  {title} {\bibinfo {title} {{Multiconfigurational
  Hartree-Fock close-coupling ansatz: Application to the argon photoionization
  cross section and delays}},\ }\href
  {https://doi.org/10.1103/PhysRevA.87.023420} {\bibfield  {journal} {\bibinfo
  {journal} {Phys. Rev. A}\ }\textbf {\bibinfo {volume} {87}},\ \bibinfo
  {pages} {023420} (\bibinfo {year} {2013})}\BibitemShut {NoStop}%
\bibitem [{\citenamefont {Argenti}\ and\ \citenamefont
  {Lindroth}(2021)}]{Argenti2021}%
  \BibitemOpen
  \bibfield  {author} {\bibinfo {author} {\bibfnamefont {L.}~\bibnamefont
  {Argenti}}\ and\ \bibinfo {author} {\bibfnamefont {E.}~\bibnamefont
  {Lindroth}},\ }\href {https://doi.org/10.48550/ARXIV.2105.10847} {\bibinfo
  {title} {Attosecond photoelectron spectroscopy of helium doubly excited
  states}} (\bibinfo {year} {2021})\BibitemShut {NoStop}%
\bibitem [{\citenamefont {Harkema}\ \emph {et~al.}(2021)\citenamefont
  {Harkema}, \citenamefont {Cariker}, \citenamefont {Lindroth}, \citenamefont
  {Argenti},\ and\ \citenamefont {Sandhu}}]{Cariker2021}%
  \BibitemOpen
  \bibfield  {author} {\bibinfo {author} {\bibfnamefont {N.}~\bibnamefont
  {Harkema}}, \bibinfo {author} {\bibfnamefont {C.}~\bibnamefont {Cariker}},
  \bibinfo {author} {\bibfnamefont {E.}~\bibnamefont {Lindroth}}, \bibinfo
  {author} {\bibfnamefont {L.}~\bibnamefont {Argenti}},\ and\ \bibinfo {author}
  {\bibfnamefont {A.}~\bibnamefont {Sandhu}},\ }\bibfield  {title} {\bibinfo
  {title} {Autoionizing polaritons in attosecond atomic ionization},\ }\href
  {https://doi.org/10.1103/PhysRevLett.127.023202} {\bibfield  {journal}
  {\bibinfo  {journal} {Phys. Rev. Lett.}\ }\textbf {\bibinfo {volume} {127}},\
  \bibinfo {pages} {023202} (\bibinfo {year} {2021})}\BibitemShut {NoStop}%
\bibitem [{\citenamefont {Argenti}\ \emph {et~al.}(2015)\citenamefont
  {Argenti}, \citenamefont {Jim\'enez-Gal\'an}, \citenamefont {Marante},
  \citenamefont {Ott}, \citenamefont {Pfeifer},\ and\ \citenamefont
  {Mart\'{\i}n}}]{LucaAlvaroPRA2015}%
  \BibitemOpen
  \bibfield  {author} {\bibinfo {author} {\bibfnamefont {L.}~\bibnamefont
  {Argenti}}, \bibinfo {author} {\bibfnamefont {A.}~\bibnamefont
  {Jim\'enez-Gal\'an}}, \bibinfo {author} {\bibfnamefont {C.}~\bibnamefont
  {Marante}}, \bibinfo {author} {\bibfnamefont {C.}~\bibnamefont {Ott}},
  \bibinfo {author} {\bibfnamefont {T.}~\bibnamefont {Pfeifer}},\ and\ \bibinfo
  {author} {\bibfnamefont {F.}~\bibnamefont {Mart\'{\i}n}},\ }\bibfield
  {title} {\bibinfo {title} {Dressing effects in the attosecond transient
  absorption spectra of doubly excited states in helium},\ }\href
  {https://doi.org/10.1103/PhysRevA.91.061403} {\bibfield  {journal} {\bibinfo
  {journal} {Phys. Rev. A}\ }\textbf {\bibinfo {volume} {91}},\ \bibinfo
  {pages} {061403} (\bibinfo {year} {2015})}\BibitemShut {NoStop}%
\bibitem [{\citenamefont {Mohr}\ \emph {et~al.}(2016)\citenamefont {Mohr},
  \citenamefont {Newell},\ and\ \citenamefont {Taylor}}]{CODdataNIST}%
  \BibitemOpen
  \bibfield  {author} {\bibinfo {author} {\bibfnamefont {P.~J.}\ \bibnamefont
  {Mohr}}, \bibinfo {author} {\bibfnamefont {D.~B.}\ \bibnamefont {Newell}},\
  and\ \bibinfo {author} {\bibfnamefont {B.~N.}\ \bibnamefont {Taylor}},\
  }\bibfield  {title} {\bibinfo {title} {Codata recommended values of the
  fundamental physical constants: 2014},\ }\href
  {https://doi.org/10.1103/RevModPhys.88.035009} {\bibfield  {journal}
  {\bibinfo  {journal} {Rev. Mod. Phys.}\ }\textbf {\bibinfo {volume} {88}},\
  \bibinfo {pages} {035009} (\bibinfo {year} {2016})}\BibitemShut {NoStop}%
\bibitem [{\citenamefont {de~Boor}(1978)}]{deBoor1978}%
  \BibitemOpen
  \bibfield  {author} {\bibinfo {author} {\bibfnamefont {C.}~\bibnamefont
  {de~Boor}},\ }\href@noop {} {\emph {\bibinfo {title} {A practical guide to
  splines}}}\ (\bibinfo  {publisher} {Springer},\ \bibinfo {year}
  {1978})\BibitemShut {NoStop}%
\bibitem [{\citenamefont {Bachau}\ \emph {et~al.}(2001)\citenamefont {Bachau},
  \citenamefont {Cormier}, \citenamefont {Decleva}, \citenamefont {Hansen},\
  and\ \citenamefont {Mart{\'{i}}n}}]{Bachau2001}%
  \BibitemOpen
  \bibfield  {author} {\bibinfo {author} {\bibfnamefont {H.}~\bibnamefont
  {Bachau}}, \bibinfo {author} {\bibfnamefont {E.}~\bibnamefont {Cormier}},
  \bibinfo {author} {\bibfnamefont {P.}~\bibnamefont {Decleva}}, \bibinfo
  {author} {\bibfnamefont {J.~E.}\ \bibnamefont {Hansen}},\ and\ \bibinfo
  {author} {\bibfnamefont {F.}~\bibnamefont {Mart{\'{i}}n}},\ }\bibfield
  {title} {\bibinfo {title} {{Applications of B-splines in atomic and molecular
  physics}},\ }\href {https://doi.org/10.1088/0034-4885/64/12/205} {\bibfield
  {journal} {\bibinfo  {journal} {Rep. Prog. Phys.}\ }\textbf {\bibinfo
  {volume} {64}},\ \bibinfo {pages} {1815} (\bibinfo {year}
  {2001})}\BibitemShut {NoStop}%
\bibitem [{\citenamefont {Rost}\ \emph {et~al.}(1997)\citenamefont {Rost},
  \citenamefont {Schulz}, \citenamefont {Domke},\ and\ \citenamefont
  {Kaindl}}]{Rost_1997}%
  \BibitemOpen
  \bibfield  {author} {\bibinfo {author} {\bibfnamefont {J.~M.}\ \bibnamefont
  {Rost}}, \bibinfo {author} {\bibfnamefont {K.}~\bibnamefont {Schulz}},
  \bibinfo {author} {\bibfnamefont {M.}~\bibnamefont {Domke}},\ and\ \bibinfo
  {author} {\bibfnamefont {G.}~\bibnamefont {Kaindl}},\ }\bibfield  {title}
  {\bibinfo {title} {Resonance parameters of photo doubly excited helium},\
  }\href {https://doi.org/10.1088/0953-4075/30/21/010} {\bibfield  {journal}
  {\bibinfo  {journal} {J. Phys. B: At. Mol. Opt. Phys.}\ }\textbf {\bibinfo
  {volume} {30}},\ \bibinfo {pages} {4663} (\bibinfo {year}
  {1997})}\BibitemShut {NoStop}%
\bibitem [{\citenamefont {B{\"{u}}rgers}\ \emph {et~al.}(1995)\citenamefont
  {B{\"{u}}rgers}, \citenamefont {Wintgen}, \citenamefont {Rost}, \citenamefont
  {Burgers},\ and\ \citenamefont {Rest}}]{Burgers1995}%
  \BibitemOpen
  \bibfield  {author} {\bibinfo {author} {\bibfnamefont {A.}~\bibnamefont
  {B{\"{u}}rgers}}, \bibinfo {author} {\bibfnamefont {D.}~\bibnamefont
  {Wintgen}}, \bibinfo {author} {\bibfnamefont {J.~M.}\ \bibnamefont {Rost}},
  \bibinfo {author} {\bibfnamefont {a.}~\bibnamefont {Burgers}},\ and\ \bibinfo
  {author} {\bibfnamefont {J.~M.}\ \bibnamefont {Rest}},\ }\bibfield  {title}
  {\bibinfo {title} {{Highly doubly excited S states of the helium atom}},\
  }\href {https://doi.org/10.1088/0953-4075/28/15/010} {\bibfield  {journal}
  {\bibinfo  {journal} {J. Phys. B: At. Mol. Opt. Phys.}\ }\textbf {\bibinfo
  {volume} {28}},\ \bibinfo {pages} {3163} (\bibinfo {year}
  {1995})}\BibitemShut {NoStop}%
\bibitem [{\citenamefont {{J{\"o}nsson}}\ \emph {et~al.}(2007)\citenamefont
  {{J{\"o}nsson}}, \citenamefont {{He}}, \citenamefont {{Froese Fischer}},\
  and\ \citenamefont {{Grant}}}]{Froese2007}%
  \BibitemOpen
  \bibfield  {author} {\bibinfo {author} {\bibfnamefont {P.}~\bibnamefont
  {{J{\"o}nsson}}}, \bibinfo {author} {\bibfnamefont {X.}~\bibnamefont {{He}}},
  \bibinfo {author} {\bibfnamefont {C.}~\bibnamefont {{Froese Fischer}}},\ and\
  \bibinfo {author} {\bibfnamefont {I.~P.}\ \bibnamefont {{Grant}}},\
  }\bibfield  {title} {\bibinfo {title} {{The grasp2K relativistic atomic
  structure package}},\ }\href {https://doi.org/10.1016/j.cpc.2007.06.002}
  {\bibfield  {journal} {\bibinfo  {journal} {Computer Physics Communications}\
  }\textbf {\bibinfo {volume} {177}},\ \bibinfo {pages} {597} (\bibinfo {year}
  {2007})}\BibitemShut {NoStop}%
\bibitem [{\citenamefont {Sakurai}\ and\ \citenamefont
  {Napolitano}(2017)}]{sakurai2017modern}%
  \BibitemOpen
  \bibfield  {author} {\bibinfo {author} {\bibfnamefont {J.}~\bibnamefont
  {Sakurai}}\ and\ \bibinfo {author} {\bibfnamefont {J.}~\bibnamefont
  {Napolitano}},\ }\href {https://books.google.com.pk/books?id=010yDwAAQBAJ}
  {\emph {\bibinfo {title} {Modern Quantum Mechanics}}}\ (\bibinfo  {publisher}
  {Cambridge University Press},\ \bibinfo {year} {2017})\BibitemShut {NoStop}%
\bibitem [{\citenamefont {Blanes}\ \emph {et~al.}(2009)\citenamefont {Blanes},
  \citenamefont {Casas}, \citenamefont {Oteo},\ and\ \citenamefont
  {Ros}}]{Blanes2009}%
  \BibitemOpen
  \bibfield  {author} {\bibinfo {author} {\bibfnamefont {S.}~\bibnamefont
  {Blanes}}, \bibinfo {author} {\bibfnamefont {F.}~\bibnamefont {Casas}},
  \bibinfo {author} {\bibfnamefont {J.}~\bibnamefont {Oteo}},\ and\ \bibinfo
  {author} {\bibfnamefont {J.}~\bibnamefont {Ros}},\ }\bibfield  {title}
  {\bibinfo {title} {The magnus expansion and some of its applications},\
  }\href {https://doi.org/https://doi.org/10.1016/j.physrep.2008.11.001}
  {\bibfield  {journal} {\bibinfo  {journal} {Physics Reports}\ }\textbf
  {\bibinfo {volume} {470}},\ \bibinfo {pages} {151} (\bibinfo {year}
  {2009})}\BibitemShut {NoStop}%
\bibitem [{\citenamefont {Argenti}\ \emph {et~al.}(2013)\citenamefont
  {Argenti}, \citenamefont {Pazourek}, \citenamefont {Feist}, \citenamefont
  {Nagele}, \citenamefont {Liertzer}, \citenamefont {Persson}, \citenamefont
  {Burgd\"orfer},\ and\ \citenamefont {Lindroth}}]{ArgentiPRA2013}%
  \BibitemOpen
  \bibfield  {author} {\bibinfo {author} {\bibfnamefont {L.}~\bibnamefont
  {Argenti}}, \bibinfo {author} {\bibfnamefont {R.}~\bibnamefont {Pazourek}},
  \bibinfo {author} {\bibfnamefont {J.}~\bibnamefont {Feist}}, \bibinfo
  {author} {\bibfnamefont {S.}~\bibnamefont {Nagele}}, \bibinfo {author}
  {\bibfnamefont {M.}~\bibnamefont {Liertzer}}, \bibinfo {author}
  {\bibfnamefont {E.}~\bibnamefont {Persson}}, \bibinfo {author} {\bibfnamefont
  {J.}~\bibnamefont {Burgd\"orfer}},\ and\ \bibinfo {author} {\bibfnamefont
  {E.}~\bibnamefont {Lindroth}},\ }\bibfield  {title} {\bibinfo {title}
  {Photoionization of helium by attosecond pulses: Extraction of spectra from
  correlated wave functions},\ }\href
  {https://doi.org/10.1103/PhysRevA.87.053405} {\bibfield  {journal} {\bibinfo
  {journal} {Phys. Rev. A}\ }\textbf {\bibinfo {volume} {87}},\ \bibinfo
  {pages} {053405} (\bibinfo {year} {2013})}\BibitemShut {NoStop}%
\bibitem [{\citenamefont {Argenti}\ and\ \citenamefont
  {Moccia}(2007)}]{Argenti2006R}%
  \BibitemOpen
  \bibfield  {author} {\bibinfo {author} {\bibfnamefont {L.}~\bibnamefont
  {Argenti}}\ and\ \bibinfo {author} {\bibfnamefont {R.}~\bibnamefont
  {Moccia}},\ }\bibfield  {title} {\bibinfo {title} {He photoionization:
  $\beta$nand $\sigma$nbelow n = 5 and 6 thresholds},\ }\href
  {https://doi.org/10.1007/s00214-007-0369-4} {\bibfield  {journal} {\bibinfo
  {journal} {Th. Chem. Acc.}\ }\textbf {\bibinfo {volume} {118}},\ \bibinfo
  {pages} {485} (\bibinfo {year} {2007})}\BibitemShut {NoStop}%
\bibitem [{\citenamefont {Fano}(1957)}]{FanoDM_1957}%
  \BibitemOpen
  \bibfield  {author} {\bibinfo {author} {\bibfnamefont {U.}~\bibnamefont
  {Fano}},\ }\bibfield  {title} {\bibinfo {title} {Description of states in
  quantum mechanics by density matrix and operator techniques},\ }\href
  {https://doi.org/10.1103/RevModPhys.29.74} {\bibfield  {journal} {\bibinfo
  {journal} {Rev. Mod. Phys.}\ }\textbf {\bibinfo {volume} {29}},\ \bibinfo
  {pages} {74} (\bibinfo {year} {1957})}\BibitemShut {NoStop}%
\bibitem [{\citenamefont {Pabst}\ \emph {et~al.}(2011)\citenamefont {Pabst},
  \citenamefont {Greenman}, \citenamefont {Ho}, \citenamefont {Mazziotti},\
  and\ \citenamefont {Santra}}]{Pabst2011}%
  \BibitemOpen
  \bibfield  {author} {\bibinfo {author} {\bibfnamefont {S.}~\bibnamefont
  {Pabst}}, \bibinfo {author} {\bibfnamefont {L.}~\bibnamefont {Greenman}},
  \bibinfo {author} {\bibfnamefont {P.~J.}\ \bibnamefont {Ho}}, \bibinfo
  {author} {\bibfnamefont {D.~A.}\ \bibnamefont {Mazziotti}},\ and\ \bibinfo
  {author} {\bibfnamefont {R.}~\bibnamefont {Santra}},\ }\bibfield  {title}
  {\bibinfo {title} {Decoherence in attosecond photoionization},\ }\href
  {https://doi.org/10.1103/PhysRevLett.106.053003} {\bibfield  {journal}
  {\bibinfo  {journal} {Phys. Rev. Lett.}\ }\textbf {\bibinfo {volume} {106}},\
  \bibinfo {pages} {053003} (\bibinfo {year} {2011})}\BibitemShut {NoStop}%
\bibitem [{\citenamefont {Mehmood}\ \emph
  {et~al.}(2021{\natexlab{b}})\citenamefont {Mehmood}, \citenamefont
  {Lindroth},\ and\ \citenamefont {Argenti}}]{Saad2021}%
  \BibitemOpen
  \bibfield  {author} {\bibinfo {author} {\bibfnamefont {S.}~\bibnamefont
  {Mehmood}}, \bibinfo {author} {\bibfnamefont {E.}~\bibnamefont {Lindroth}},\
  and\ \bibinfo {author} {\bibfnamefont {L.}~\bibnamefont {Argenti}},\
  }\bibfield  {title} {\bibinfo {title} {Coherence control in helium-ion
  ensembles},\ }\href {https://doi.org/10.1103/PhysRevResearch.3.023233}
  {\bibfield  {journal} {\bibinfo  {journal} {Phys. Rev. Research}\ }\textbf
  {\bibinfo {volume} {3}},\ \bibinfo {pages} {023233} (\bibinfo {year}
  {2021}{\natexlab{b}})}\BibitemShut {NoStop}%
\end{thebibliography}
\end{document}